\documentclass[11pt]{article}
\usepackage{amssymb,amsmath}
\usepackage{epsfig}
\usepackage{hyperref}
\usepackage{cite} % to automatically sort the cite numbers

\newcommand{\qedsymb}{\hfill{\rule{2mm}{2mm}}}
\newenvironment{proof}[1][]{\begin{trivlist}
\item[\hspace{\labelsep}{\bf\noindent Proof#1:\/}] }{\qedsymb\end{trivlist}}

\mathchardef\ordinarycolon\mathcode`\:     % WVD: this is for a correctly aligned :=
\mathcode`\:=\string"8000
\def\vcentcolon{\mathrel{\mathop\ordinarycolon}} \begingroup
\catcode`\:=\active \lowercase{\endgroup \let :\vcentcolon }

\newcommand{\ignore}[1]{}

\newtheorem{theorem}{Theorem}[section]

\newtheorem{deff}[theorem]{Definition}

\newtheorem{claim}[theorem]{Claim}
\newtheorem{lemma}[theorem]{Lemma}

\newtheorem{coro}[theorem]{Corollary}
\newtheorem{fact}[theorem]{Fact}

\newenvironment{remark}[1][] {
  \smallbreak \noindent {\bf Remark#1:~}} {
  \par\medbreak
}

\newcommand{\poly}{{\mathrm{poly}}}
\newcommand{\smfrac}[2]{\mbox{$\frac{#1}{#2}$}}

\newcommand{\ket}[1]{|#1\rangle}
\newcommand{\bra}[1]{\langle#1|}
\newcommand{\ontop}[2]{{\begin{array}{l} {#1} \\ {#2} \end{array}}}
\newcommand{\stackket}[2]{{\Big | \hskip -5pt \ontop{#1}{#2} \hskip -3pt  \Big \rangle}}
\newcommand{\stackbra}[2]{{\Big \langle \hskip -5pt \ontop{#1}{#2} \hskip -3pt  \Big |}}
\newcommand{\stackketbra}[2]{{ \stackket{#1}{#2} \stackbra{#1}{#2} }}
\newcommand{\ra}{\rangle}
\newcommand{\ketbra}[2]{|#1\rangle\langle#2|}

\newcommand{\eps}{\epsilon}

\def\final{{\mathrm{final}}}
\def\init{{\mathrm{init}}}
\def\clock{{\mathrm{clock}}}
\def\clockinit{{\mathrm{clockinit}}}
\def\rminput{{\mathrm{input}}}
\def\rules{{\mathrm{rules}}}

\newcommand{\dnote}[1]{}
\newcommand{\onote}[1]{}
\newcommand{\jnote}[1]{}
\newcommand{\wnote}[1]{}

\def\calH{{\cal{H}}}
\def\calS{{\cal{S}}}

\def\NP{{\sf{NP}}}

\newlength{\bigcirclen}
\def\mns{{\mbox{-}}}

\def\statea{{\bigcirc}}
\def\stateba{{\settowidth{\bigcirclen}{$\bigcirc$}\makebox[0pt][l]{\makebox[\bigcirclen][c]{$\uparrow$}}\bigcirc}}
\def\statebb{{\settowidth{\bigcirclen}{$\bigcirc$}\makebox[0pt][l]{\makebox[\bigcirclen][c]{$\downarrow$}}\bigcirc}}
\def\stateb{{\settowidth{\bigcirclen}{$\bigcirc$}\makebox[0pt][l]{\makebox[\bigcirclen][c]{$\uparrow$}}\makebox[0pt][l]{\makebox[\bigcirclen][c]{$\downarrow$}}\bigcirc}}
\def\stateca{{\settowidth{\bigcirclen}{$\bigcirc$}\makebox[0pt][l]{\makebox[\bigcirclen][c]{$\Uparrow$}}\bigcirc}}
\def\statecb{{\settowidth{\bigcirclen}{$\bigcirc$}\makebox[0pt][l]{\makebox[\bigcirclen][c]{$\Downarrow$}}\bigcirc}}
\def\statec{{\settowidth{\bigcirclen}{$\bigcirc$}\makebox[0pt][l]{\makebox[\bigcirclen][c]{\raisebox{.035em}{\fontsize{9}{0}\selectfont$\Updownarrow$}}}\bigcirc}}
\def\stated{{\settowidth{\bigcirclen}{$\bigcirc$}%
\makebox[0pt][l]{\makebox[\bigcirclen][c]{$\times$}}\bigcirc}}

\title{Adiabatic Quantum Computation is Equivalent to Standard \\ Quantum Computation}

\author{
Dorit Aharonov \\ School of Computer Science and Engineering,\\
Hebrew University,
Jerusalem,
 Israel
\and
Wim van Dam\\Department of Computer Science,\\ UC Santa Barbara, CA
\and
Julia Kempe\\CNRS-LRI UMR 8623,\\ Universit\'e de Paris-Sud, Orsay, France
\and
Zeph Landau\\Department of Mathematics,\\ City College of New York, NY
\and
Seth Lloyd\\Department of Mechanical Engineering,\\ MIT, Cambridge, MA
\and
Oded Regev\\Computer Science Department,\\ Tel Aviv University,
Israel }
\ignore{
\author{~~~~~~~~~~~~~~~~~~~~~~~~~~~~~~~Dorit
Aharonov\thanks{School of Computer Science and Engineering,
Hebrew University, % Jerusalem, 91904,
Israel.
Also Computer Science Division, UC Berkeley, CA 94720.}
\and
Wim van Dam\thanks{Department of Physics, MIT, Cambridge, MA}
\and \newline
Julia Kempe\thanks{CNRS-LRI UMR 8623, Universit\'e de Paris-Sud,Orsay,
France.  Also Computer Science Division and Department of Chemistry,
UC Berkeley, CA 94720} %91405
\and \and\and\and\and\and
\and\and\and\and\and
Zeph Landau\thanks{Department of Mathematics, City College of New York, NY.}
\and
Seth Lloyd\thanks{Department of Mechanical Engineering,MIT, Cambridge, MA.}
\and
Oded Regev\thanks{Computer Science Department, Tel-Aviv University,
Israel}}}

\begin{document}

\maketitle

\begin{abstract}
Adiabatic quantum computation
 has recently attracted attention in the physics and computer science
communities, but its computational power was unknown.
We describe an efficient adiabatic simulation
 of any given quantum algorithm, which implies that
 the adiabatic computation model and the conventional quantum
computation model are polynomially equivalent.
Our result can be extended to the
physically realistic setting of
particles arranged on a two-dimensional grid with nearest neighbor
interactions. The equivalence between the models
provides a new vantage point from which to tackle
the central issues in quantum computation, namely
designing new quantum algorithms and
constructing fault tolerant quantum computers.
In particular, by translating the main open questions
in the area of quantum algorithms to the language of spectral gaps
of sparse matrices, the result
makes these questions accessible to a
wider scientific audience, acquainted with
mathematical physics, expander theory and rapidly mixing Markov chains.
\ignore{
The model of adiabatic quantum computation
 has recently attracted attention in the physics and computer science
communities, but its exact computational power has been unknown.
We settle this question and describe an efficient adiabatic simulation
 of any given quantum algorithm. This implies that
 the adiabatic computation model and the standard quantum
circuit model are polynomially equivalent.
We also describe an extension of this result with implications to
physical implementations of adiabatic computation.
We believe that our result highlights the potential importance
of the adiabatic computation model in the design of quantum algorithms
and in their experimental realization.}
\end{abstract}

\section{Introduction}
The study of adiabatic quantum computation was initiated several years
ago by Farhi, Goldstone, Gutmann and Sipser \cite{farhiad}, who
suggested a novel quantum algorithm for solving classical optimization
problems such as {\sc Satisfiability} ({\sc Sat}).  Their algorithm is
based on a celebrated theorem in quantum mechanics known as {\em the
adiabatic theorem} \cite{kato,messiah}. Although the exact worst-case
behavior of this algorithm is not known, several simulations (see,
e.g., \cite{farhi}) on random instances of up to $20$ quantum bits led
to various optimistic speculations. The bad news is that there is now
mounting evidence \cite{vandamvaz, vandam2, ben} that the algorithm of
\cite{farhiad} takes exponential time in the worst-case for
$\NP$-complete problems.  Nevertheless, adiabatic computation was
since shown to be promising in other less ambitious directions: it
possesses several interesting algorithmic capabilities, as we will
soon review, and in addition, it exhibits inherent robustness against
certain types of quantum errors \cite{preskill}. We note that a small
scale adiabatic algorithm has already been implemented experimentally,
using a Nuclear Magnetic Resonance (NMR) system \cite{adnmr}.

We briefly describe the model of adiabatic computation (a more precise
description appears in Section~\ref{sec:model}).  A computation in
this model is specified by two Hamiltonians named $H_\init$ and
$H_\final$ (a Hamiltonian is simply a Hermitian matrix).  The
eigenvector with smallest eigenvalue (also known as the \emph{ground
state}) of $H_\init$ is required to be an easy to prepare state, such
as a tensor product state.  The output of the adiabatic computation is
the ground state of the final Hamiltonian $H_\final$.  Hence, we
choose an $H_\final$ whose ground state represents the solution to our
problem.  We require the Hamiltonians to be \emph{local}, i.e., we
require them to only involve interactions between a constant number of
particles (this can be seen as the equivalent of allowing gates
operating on a constant number of qubits in the standard model).
This, in particular, makes sure that the Hamiltonians have a short
classical description, by simply listing the matrix entries of each
local term.
\ignore{We also require the eigenvalues of the Hamiltonians to be
bounded by a polynomial in the resources.}  The running time of the
adiabatic computation is determined by the minimal spectral
gap\footnote{The spectral gap is the difference between the lowest and
second lowest eigenvalue.}  of all the Hamiltonians on the straight
line connecting $H_\init$ and $H_\final$: $H(s)=(1-s)H_\init + s
H_\final$ for $s \in [0,1]$.  More precisely, the adiabatic
computation is polynomial time if this minimal spectral gap is at
least inverse polynomial.

The motivation for the above definition is physical.  The Hamiltonian
operator corresponds to the energy of the quantum system, and for it
to be physically realistic and implementable it must be local. Its
ground state is the state of lowest energy. We can set up a quantum
system in the ground state of $H_\init$ (which is supposed to be easy
to generate) and apply the Hamiltonian $H_\init$ to the system.  We
then slowly modify the Hamiltonian along the straight line from
$H_\init$ towards $H_\final$. It follows from the adiabatic theorem
that if this transformation is performed slowly enough (how slow is
determined by the minimal spectral gap), the final state of the system
will be in the ground state of $H_\final$, as required.

What is the computational power of this model?  In order to refer to
the adiabatic model as a computational model that computes classical
functions (rather than quantum states), we consider the result of the
adiabatic computation to be the outcome of a measurement of one or
more of the qubits, performed on the final ground state. It is known that
adiabatic computation can be efficiently simulated by standard quantum
computers \cite{vandamvaz,farhi}.  Hence, its computational power is
not greater than that of standard quantum computers.  Several positive
results are also known. In \cite{vandamvaz,cerf} it was shown that
Grover's quadratic speed-up for an unsorted search \cite{groversearch}
can be realized as an adiabatic computation. Moreover,
\cite{farhianneal,ben,santoro} showed that adiabatic computation can
`tunnel' through wide energy barriers and thus outperform simulated
annealing, a classical counterpart of the adiabatic model.  However,
whether adiabatic computation can achieve the full power of quantum
computation was not known.  In fact, even the question of whether
adiabatic computation can simulate general \emph{classical}
computations efficiently was unknown. The focus of this paper is the
exact characterization of the computational power of adiabatic
computation.

Before we describe our results, let us clarify one subtle point.  Most
of the previous work on the subject focused on a restricted class of
adiabatic algorithms that can be referred to as adiabatic {\it
optimization} algorithms. In these algorithms, $H_\final$ is chosen to
be a diagonal matrix, corresponding to a combinatorial optimization
problem. In particular, this implies that the ground state of
$H_\final$ (which is the output of the computation) is a classical
state, i.e., a state in the computational basis.  In this paper,
however, we associate the term \emph{adiabatic computation} with the
more general class of adiabatic algorithms, where the only restriction
on $H_\final$ is that it is a local Hamiltonian.  We do this because,
from a physical point of view, there is no reason to force the
physical process described above to have a diagonal $H_\final$, when
all other Hamiltonians are not restricted this way. Thus, our
definition of adiabatic computation seems to be the natural one to
use. It is this natural definition that allows us to prove our
results.

\subsection{Results -- Computational Complexity of the Adiabatic Model}
Our main result clarifies the question of the computational power of
adiabatic algorithms.  We show:
\begin{theorem}\label{thm:main}
 The model of adiabatic computation is polynomially equivalent to the
 standard model of quantum computation.
\end{theorem}

As mentioned above, one direction of the equivalence is already known
 \cite{farhi,vandamvaz}. Our contribution is to show that standard
 quantum computation can be efficiently simulated by adiabatic
 computation. We do this by using adiabatic computation with $3$-local
 Hamiltonians.  We note that \cite{amnon} made a preliminary step in
 the direction of Theorem~\ref{thm:main} but the model that they
 considered was quite different.\footnote{Namely, \cite{amnon} showed
 that adiabatic computation using \emph{simulatable} Hamiltonians is
 as powerful as standard quantum computation. Simulatable Hamiltonians
 are Hamiltonians that can be simulated efficiently by a quantum
 circuit.  They are very different from local Hamiltonians, and they
 cannot even be written explicitly. Instead, such Hamiltonians are
 specified using products of local unitary matrices.}

One corollary of our main theorem is the following. We can consider
the model of adiabatic computation with a more general set of
Hamiltonians known as \emph{explicit sparse} Hamiltonians. These are
Hermitian matrices that have at most polynomially many nonzero
elements in each row and column, and,
moreover, there is an efficient Turing machine that can generate a
list of all nonzero entries in a given row or column. Clearly, local Hamiltonians are a special case of
explicit sparse Hamiltonians.  It was shown in \cite{amnon} that
adiabatic computation with explicit sparse Hamiltonians can still be
simulated by standard quantum computation (this extends the result of
\cite{vandamvaz,farhiad} in a non-trivial way). Hence, we obtain
the following result.
\begin{coro}\label{corol:main}
 The model of adiabatic computation with explicit sparse Hamiltonians
 is polynomially equivalent to the standard model of quantum
 computation.
\end{coro}
Explicit sparse matrices are pervasive in computer science and
combinatorics, and hence this corollary might be more useful than
Theorem~\ref{thm:main} in the context of the design of quantum
algorithms and the study of quantum complexity.

To summarize, our results show that questions about quantum
computation can be equivalently considered in the model of adiabatic
computation, a model that is quite different from the more common
circuit-based models.  There are two reasons why it seems worthwhile
to try to design quantum algorithms in the adiabatic framework.
First, there are several known powerful techniques to analyze spectral
gaps of matrices, including expander theory \cite{expanders} and
rapidly mixing Markov chains \cite{lovasz,sinclair}.  Indeed,
probability theory is often used in mathematical physics to analyze
spectral gaps of Hamiltonians (see, e.g., \cite{spitzer}), and our
proofs also make extensive use of Markov chain tools.  Second, it is
known that many interesting algorithmic problems in quantum
computation can be cast as quantum state generation problems
\cite{amnon}.  The problem of generating special quantum states seems
more natural in the adiabatic model than in the standard model.

\subsection{Results -- Towards Experimental Implications}
Theorem~\ref{thm:main} uses 3-local Hamiltonians that act on
particles that may be arbitrarily far apart.
{}From a practical point of view,
it is often difficult to create controlled interactions between
particles located far-away from each other. Moreover,
three-particle Hamiltonians
are technologically very difficult to realize.
If one wants to physically realize the adiabatic algorithms,
it would be much better to have only
two-local interactions between nearest neighbor particles.
 To this end
 we prove the following theorem.  This, we believe, brings the
adiabatic computation model one step closer to physical realization.
\begin{theorem}\label{thm:geo}
Any quantum computation can be efficiently simulated by an adiabatic
computation with two-local nearest neighbor Hamiltonians operating on
six-state particles set on a two dimensional grid.
\end{theorem}
The need for six-state particles arises from our construction.  It is
an open question whether this can be improved.

Theorems~\ref{thm:main} and~\ref{thm:geo} open up the possibility of
physically realizing universal quantum computation
using adiabatically evolving quantum systems.
As mentioned before, there are possible advantages to this approach:
adiabatic quantum computation is resilient to certain types
of noise \cite{preskill}. An important component of this
resilience is the existence of a spectral gap in the Hamiltonian.
It is well known in physics that such a gap
plays an important role in the context of protecting quantum systems from
noise. However, it remains to be
further studied, both experimentally and
theoretically, what the right model for noisy adiabatic computation is,
and whether fault tolerant adiabatic computation can be achieved.
We refer the reader to further discussion in Subsection \ref{sec:open}.

\subsection{Proof of Theorem~\ref{thm:main}: Overview}
Given an arbitrary quantum circuit \cite{nielsen}, our goal is to
design an adiabatic computation whose output is the same as that of
the quantum circuit.  Some similarities between the models are
obvious: one model involves unitary gates on a constant number of
qubits, while the other involves local Hamiltonians.  However, after
some thought, one eventually arrives at the following difficulty.  The
output state of the adiabatic computation is the ground state of
 $H_\final$.  The
output state of the quantum circuit is its final state, which is
unknown to us. How can we specify $H_\final$ without knowing the
output state of the quantum circuit?  Notice that this state can be
some complicated quantum superposition. One might wonder why our task
is not trivial, since this state does have an efficient local
classical description, namely the quantum circuit.  However, local
quantum gates, which operate in sequence to generate a non-local
overall action, are very different from local Hamiltonians, which
correspond to simultaneous local constraints.  To explain the
solution, we first set some notations.

Without loss of generality we assume that the input to the quantum
circuit consists of $n$ qubits all initialized to
$\ket{0}$'s.\footnote{Otherwise, the first $n$ gates can be used to
flip the qubits to the desired input.} Then, a sequence of $L$ unitary
gates, $U_1,\dots,U_L$, each operating on one or two qubits, is
applied to the state. The system's state after the $\ell$'th gate is
$\ket{\alpha(\ell)}$.  The output of the quantum circuit is in general
a complicated quantum state $\ket{\alpha(L)}$ of $n$ qubits, which is
then measured in the standard basis.  We now want to associate with it
a corresponding adiabatic computation.

A first natural attempt would be to define $H_{\final}$ as a local
Hamiltonian with $\ket{\alpha(L)}$ as its ground state.  However, this
attempt encounters the difficulty mentioned above: not knowing
$\ket{\alpha(L)}$, it seems impossible to explicitly specify
$H_\final$. The key to resolve this difficulty is the observation that
the ground state of $H_\final$ need not necessarily be the state
$\ket{\alpha(L)}$. It is sufficient (under some mild restrictions)
that the ground state has a non-negligible inner product with
$\ket{\alpha(L)}$. This gives us significant flexibility in designing
$H_\final$. Our idea is to base our solution on a seemingly unrelated
ingenious result of Kitaev \cite{Kitaev:book}, in which he provides
the first quantum $\NP$-complete problem, namely, {\sc local
Hamiltonians}.  This result can be viewed as the quantum analogue of
the Cook-Levin theorem \cite{papa}, which states that $3$-{\sc Sat} is
$\NP$-complete.  For his proof, Kitaev defined a local Hamiltonian
that checks the time propagation of a quantum circuit.  Kitaev's local
Hamiltonian has as its ground state the entire \emph{history} of the
quantum computation, in \emph{superposition}:
\begin{eqnarray}\label{eq:final_state}
\ket{\eta} &:=& \frac{1}{\sqrt{L+1}} \sum_{\ell=0}^L
\ket{\alpha(\ell)} \otimes \ket{1^\ell 0^{L-\ell}}^c.
\end{eqnarray}
The right ($L$ qubits) register is a clock that counts the steps by
adding $1$s from left to right.  The superscript $c$ denotes clock
qubits.  We note that this state has a non-negligible projection on
our desired state $\ket{\alpha(L)}$.  Hence, instead of designing a
Hamiltonian that has the final unknown state of the circuit as its
ground state, a task that seems impossible, we can define $H_\final$
to be Kitaev's local Hamiltonian. Why is it possible to define a local
Hamiltonian whose ground state is $\ket{\eta}$, whereas the same task
seems impossible with $\ket{\alpha(L)}$?  The idea is that the unary
representation of the clock enables a local verification of correct
propagation of the computation from one step to the next, which cannot
be done without the intermediate computational steps.

We thus choose Kitaev's Hamiltonian \cite{Kitaev:book} to be our
$H_\final$.  This Hamiltonian involves five body interactions (three
clock particles and two computation particles).  For the initial
Hamiltonian $H_\init$ we require that it has $\ket{\alpha(0)} \otimes \ket{0^{L}}^c$, the
first term in the history state, as its unique ground state. It is
easy to define such a local Hamiltonian, because $\ket{\alpha(0)} \otimes \ket{0^{L}}^c$ is a
tensor product state. Crucially, $H_\init$ and $H_\final$ can be
constructed efficiently from the given quantum circuit;
no knowledge of $\ket{\alpha(L)}$ is required for the construction.

A technical problem lies in showing that the spectral gap of the
intermediate Hamiltonian $H(s)$ is lower bounded by some inverse
polynomial (more specifically, we show it is larger than $1/L^2$).  To
do this, we use a mapping of the Hamiltonian to a Markov chain
corresponding to a random walk on the $L+1$ time steps. We then apply
the conductance bound from the theory of rapidly mixing Markov chains
\cite{sinclair} to bound the spectral gap of this chain.  We note
that, in general, applying the conductance bound requires knowing the
limiting distribution of the chain, which in our case is hard since it
corresponds to knowing the coefficients of the ground state for
all the Hamiltonians $H(s)$.
 We circumvent this problem by noticing that it is actually
sufficient in our case to know very little about the limiting
distribution of the Markov chain, namely that it is monotone (in a
certain sense to be defined later).  This allows us to apply the
conductance bound, and deduce that the spectral gap is
$\Omega(1/L^2)$.  From this it follows that the running time of the
adiabatic computation is polynomial.
Extracting the output of the quantum circuit from the history state
efficiently is easy: Measure all the qubits of the clock and if the
clock is in the state $\ket{1^\ell}$, the computational qubits carry
the result of the circuit.  Otherwise, start from
scratch.\footnote{This gives an overhead factor of $L$ which can be avoided
by adding $O(\frac{1}{\eps}L)$ identity gates to the quantum
circuit at the end, which has the effect that most of the history state
$|\eta\ra$ is concentrated on the final state $\ket{\alpha(L)}$. See
Subsection~\ref{sec:nonsense} for more details.}

\ignore{This scheme would not suffice to prove Theorem~\ref{thm:geo}.
The basic problem lies in arranging sufficient interaction between the
computational and the clock particles, since if the particles are set
on a grid, each clock particle can only interact with four neighbors.
We circumvent this problem as follows. Instead of having separate
clock and computational particles, we now assign to each particle both
clock and computational degrees of freedom (this is what makes our
particles six-state).  We then construct a computation that propagates
locally over the entire set of particles, snaking up and down each
column of the lattice.  The adiabatic evolution would now end up in
the history state of this snake-like sequence of states.}

The above scheme gives a proof of Theorem~\ref{thm:main} that uses
$5$-local Hamiltonians, and runs in time roughly $O(L^5)$.
The improvement to $3$-locality is based on a
simple idea (used in \cite{Kempe:03a} to prove that the $3$-local
Hamiltonian problem is quantum $\NP$-complete) but obtaining a lower
bound on the spectral gap is significantly more involved technically.
We postpone its explanation to the body of the paper.  The running
time we achieve in this case is roughly $O(L^{14})$.
\dnote{verify the numbers}

\subsection{Proof of Theorem~\ref{thm:geo}: Overview}
The idea underlying the proof of Theorem~\ref{thm:main} by itself does
not suffice to prove Theorem~\ref{thm:geo}.  The basic problem lies in
arranging sufficient interaction between the computational and the
clock particles, since if the particles are set on a grid, each clock
particle can only interact with four neighbors.  We circumvent this
problem as follows. Instead of having separate clock and computational
particles, we now assign to each particle both clock and computational
degrees of freedom (this is what makes our particles six-state).  We
then construct a computation that propagates locally over the entire
set of particles, snaking up and down each column of the lattice.  The
adiabatic evolution now ends up in the history state of this
snake-like sequence of states.

The lower bound on the spectral gap is obtained in an essentially
identical way as in the $3$-local Hamiltonian case.

\subsection{Related Work}\label{ssec:related}
 After the preliminary version of this paper appeared \cite{thisfocs},
 the results regarding Quantum-$\NP$ completeness were tightened by
 \cite{KKR:04} to show that the $2$-local Hamiltonian problem is
 Quantum-$\NP$ complete.  Following the ideas presented in the current
 paper, \cite{KKR:04} used their result to show that
Theorem~\ref{thm:main} holds when the Hamiltonians are $2$-local.

The idea to use an inverse polynomial spectral gap for fault
tolerance is certainly not new. It is a crucial ingredient in
topological (and later, geometrical) quantum computation
\cite{geometric,anyons,holonomic}.
Note, however, that in those models the
spectral gap has no effect on the running time or on any other
algorithmic aspects, and it is used only to separate the computational
subspace from the ``noisy'' subspace. In contrast, the spectral gap in
adiabatic computation is crucial from the algorithmic point of view,
since it determines the time complexity of the computation.

\subsection{Open Questions}\label{sec:open}
This paper demonstrates that quantum computation can be studied and
implemented entirely within the adiabatic computation model, without
losing its computational power. This result raises many open questions
in various directions. First, it would be interesting to determine if
the parameters presented in this work can be improved.  For example,
it might be possible to shorten the running time of our adiabatic
simulation.  Decreasing the dimensionality of the particles used in
Theorem~\ref{thm:geo} from six to two or three might be important for
implementation applications. An interesting question is whether
Theorem~\ref{thm:geo} can be achieved using a one dimensional instead
of a two dimensional grid.

Second, the possibility of fault tolerant adiabatic computation
deserves to be studied both experimentally and theoretically.  Since
the publication of the preliminary version of the current paper
\cite{thisfocs}, several researchers have begun to study adiabatic
computation in the presence of noise \cite{aberg,cerf2, lidar}.
However, it is still unclear whether adiabatic evolution
might be helpful for the physical implementation of quantum computers.

Our results imply the equivalence between standard quantum computation
and various other variants of adiabatic computation that have been
considered in the literature and that are more general than our model.
These include adiabatic computation with a general path between
$H_\init$ and $H_\final$, rather than a straight line (see
\cite{amnon} and \cite{farhipaths} for a rigorous definition), and
adiabatic computation with explicit sparse Hamiltonians \cite{amnon}
(see Corollary~\ref{corol:main}). A problem we leave open is to
characterize the computational power of adiabatic optimization,
studied in \cite{vandamvaz, vandam2,farhiad}. In this model, the
initial state is a tensor product of qubits in the state
$\frac{1}{\sqrt{2}}(|0\ra+|1\ra)$, the final Hamiltonian is diagonal,
and the evolution is carried out on a straight line. It is still
possible that such a computation can be simulated efficiently by a
classical Turing Machine.

Finally, we hope that the adiabatic framework might lead to the
discovery of new quantum algorithms. As shown in this paper, as well
as in \cite{amnon}, tools from probability theory, mathematical
physics and spectral gap analysis might turn out to be relevant and
useful.  In order to improve our understanding of the benefits of the
adiabatic paradigm, it might be insightful to see adiabatic versions
of known quantum algorithms, presented in a meaningful way.

\paragraph{Organization:}
In Section~\ref{sec:prelim} we describe the model of adiabatic
computation and state some relevant facts about Markov chains.
Section~\ref{sec:nogeo} shows how adiabatic systems with local
Hamiltonians allowing five- and later three-body interactions, can
efficiently simulate standard quantum computations.  Section
\ref{sec:geometric} shows how to adapt the construction to a
two-dimensional grid.

\section{Preliminaries}\label{sec:prelim}
\subsection{Hamiltonians of $n$-Particle Systems}
For background on $n$-qubit systems, quantum circuits and
Hamiltonians, see \cite{nielsen}. An $n$-particle system is
described by a state in Hilbert space of dimension $d^n$,
the tensor product of $n$ $d$-dimensional Hilbert spaces.
For simplicity, we restrict our discussion in this subsection
to quantum systems composed of $2$-dimensional particles, i.e., qubits;
a similar discussion holds for higher dimensional particles
(such as the $6$-dimensional case we consider later).

In the standard model of quantum computation, the state of $n$ qubits
evolves in discrete time steps by unitary operations.  In fact, the
underlying physical description of this evolution is continuous, and
is governed by Schr{\"o}dinger's equation: \( -i
\frac{d}{dt}\ket{\psi(t)}=H(t)\ket{\psi(t)}\). Here $|\psi(t)\ra$ is
the state of the $n$ qubits at time $t$, and $H(t)$ is a Hermitian
$2^n\times 2^n$ matrix operating on the space of $n$ qubits.  This
$H(t)$ is the \emph{Hamiltonian} operating on a system; it governs the
dynamics of the system. Given that the state of the system at time
$t=0$ is equal to $|\psi(0)\ra$, one can in principle solve
Schr{\"o}dinger's equation with this initial condition, to get
$|\psi(T)\ra$, the state of the system at a later time $t=T$.  The
fact that the Hamiltonian is Hermitian corresponds to the familiar
fact that the discrete time evolution of the quantum state from time
$t_1$ to a later time $t_2$ is unitary.

We sometimes refer to eigenvalues of Hamiltonians as {\em energies}.
The {\em ground energy} of a Hamiltonian is its lowest eigenvalue and
the corresponding eigenvector(s) are called {\em ground state}(s).
We define $\Delta(H)$, the \emph{spectral gap} of a
Hamiltonian $H$, to be the difference between the lowest eigenvalue of
$H$ and its second lowest eigenvalue.
 ($\Delta(H)=0$ if the lowest eigenvalue is degenerate, namely, has more than
one eigenvector associated with it).
We define the \emph{restriction} of $H$ to some subspace $\calS$, denoted
$H_\calS$, as $\Pi_\calS H \Pi_\calS$ where $\Pi_\calS$ is the orthogonal
projection on $\calS$.

 A Hamiltonian on an $n$-particle system represents a certain physical
operation that one can, in principle, apply to an $n$-particle system.
However, it is clear that one cannot efficiently apply any arbitrary
Hamiltonian (just describing it requires roughly $2^{2n}$ space).  We say that a
Hamiltonian $H$ is {\em $k$-local} if $H$ can be written as $\sum_A
H^A$ where $A$ runs over all subsets of $k$ particles, and $H^A$
operates trivially on all but the particles in $A$ (i.e., it is a
tensor product of a Hamiltonian on $A$ with identity on the particles
outside of $A$).  Notice that for any constant $k$, a $k$-local
Hamiltonian on $n$-qubits can be described by $2^{2k}n^k=\poly(n)$
numbers. We say that $H$ is local if $H$ is $k$-local for some
constant $k$.

In this paper we restrict our attention to $k$-local Hamiltonians.
This requirement corresponds to the fact that all known interactions
in nature involve a constant number of particles.  We attempt to make
$k$ as small as possible to make the Hamiltonian easier to implement.

\subsection{The Model of Adiabatic Computation}\label{sec:model}

The cornerstone of the adiabatic model of computation is the
celebrated adiabatic theorem \cite{kato,messiah}.
Consider a time-dependent Hamiltonian $H(s)$,
$s\in [0,1]$, and a system initialized at time $t=0$ in the ground
state of $H(0)$ (here and in the following we assume that for all $s\in [0,1]$, $H(s)$ has a
unique ground state). Let the system evolve according to the Hamiltonian
$H(t/T)$ from time $t=0$ to time $T$. We refer to such
a process as an \emph{adiabatic evolution according to $H$ for time $T$}.
The adiabatic theorem affirms that for large enough $T$ the
final state of the system is very close to the ground state of $H(1)$.
Just how large $T$ should be for this to happen is determined by the spectral gap
of the Hamiltonians $H(s)$. Such an upper bound on $T$ is given in the following theorem,
adapted from \cite{ben} (whose proof in turn is based on \cite{avron};
see also \cite{odedandris} for a recent elementary proof of a slightly
weaker version).

\begin{theorem}[The Adiabatic Theorem (adapted from \cite{ben})]\label{thm:ad}
Let $H_{\init}$ and $H_{\final}$ be two
Hamiltonians acting on a quantum system and consider the time-dependent
Hamiltonian $H(s):=(1-s)H_\init + s H_\final$.
Assume that for all $s$, $H(s)$ has a unique ground state.
Then for any fixed $\delta > 0$, if
\begin{eqnarray}\label{eq:adiabatic_cond}
T &\ge& \Omega
 \left(\frac{\|H_{\final}-H_{\init}\|^{1+\delta}}{\epsilon^{\delta}\min_{s
 \in [0,1]}\{\Delta^{2+\delta}(H(s))\}} \right)
\end{eqnarray}
then the final state of an adiabatic evolution according to $H$ for
time $T$ (with an appropriate setting of global phase)
is $\eps$-close in $\ell_2$-norm to the ground state of $H_\final$.
The matrix norm is the spectral norm $\|H\| := \max_w \|Hw\|/\|w\|$.
\end{theorem}
One should think of $\delta$ as being some
fixed constant, say, $0.1$. We cannot
take $\delta=0$ because of the constant hidden in the $\Omega$ notation,
which goes to infinity as $\delta$ goes to $0$.

Let us now describe the model of adiabatic computation.
In this paper we use the following definition of adiabatic computation
that slightly generalizes that of Farhi et al. \cite{farhiad}.
The adiabatic `circuit'
is determined by $H_\init$ and $H_\final$ and the output of
the computation is (close to) the ground state of $H_\final$.

\begin{deff}\label{def:ad}
A $k$-local adiabatic computation $AC(n,d,H_\init,H_\final, \eps)$ is
specified by two $k$-local Hamiltonians, $H_{\init}$ and $H_{\final}$
acting on $n$ $d$-dimensional particles,
% such that $\|H_\init\|,\|H_\final\| \leq 1$, and
 such that both Hamiltonians have
unique ground states.  The ground state of $H_{\init}$ is a tensor
product state. The output is a state that is $\eps$-close in
$\ell_2$-norm to the ground state of $H_\final$.  Let $T$
be the smallest time such that the final
state of an adiabatic evolution according to $H(s):=(1-s)
H_\init + s H_\final$ for time $T$ is $\epsilon$-close in $\ell_2$-norm
to the ground state of $H_{\final}$. The running time of the adiabatic
algorithm is defined to be $T\cdot \max_s\|H(s)\|$.
\end{deff}

Observe that we have chosen our definition of running time to  be
$T\cdot\max_s\|H(s)\|$ and not $T$.
Notice that if the Hamiltonians are multiplied by some factor, this
divides the bound of Equation \ref{eq:adiabatic_cond}, and hence $T$,
by the same factor. Hence, if the running time is defined to be $T$
one would be able to achieve arbitrarily
 small running times, by multiplying the Hamiltonians by large factors.
Our definition, on the other hand, is invariant under a multiplication
of the Hamiltonian by
an overall factor, and so takes into account the known physical
trade-off between time and energy.\footnote{
This trade-off between time and the norm of
the Hamiltonian (namely, the energy), is manifested
in Schr{\"o}dinger's equation whose solution does not change
if time is divided by some factor and at the same time
the Hamiltonian is multiplied by the same factor.}

The right hand side of Equation~\ref{eq:adiabatic_cond} can be
 used to provide an upper bound on the running time of an adiabatic
 computation.
Hence, in order to show that an adiabatic algorithm is
efficient, it is enough to use Hamiltonians of at most $\poly(n)$ norm,
and show that  for all $s\in [0,1]$ the spectral
gap $\Delta(H(s))$ is at least inverse polynomial in $n$.

We note that in certain cases, it is possible to obtain a stronger
upper bound on the running time. Indeed, assume there exists a subspace
$\calS$ such that for all $s\in [0,1]$, $H(s)$ leaves $\calS$ invariant,
i.e., $H(s)(\calS)\subseteq \calS$. Equivalently, $H(s)$ is block diagonal
in $\calS$ and its orthogonal space $\calS^\perp$. Consider $H_\calS(s)$,
the restriction of $H(s)$ to $\calS$. Then, starting from a state inside
$\calS$, an adiabatic evolution according to $H$ is {\em identical} to
an adiabatic evolution according to $H_\calS$ (this follows from
Schr{\"o}dinger's equation). Hence, we
can potentially obtain a stronger upper bound by replacing
$\Delta(H(s))$ with $\Delta(H_\calS(s))$ in Equation~\ref{eq:adiabatic_cond}.
This stronger upper bound will be used in our first adiabatic algorithm.

Finally, let us mention that one can define more general models
of adiabatic computation. For example, one might consider non-local
Hamiltonians (see \cite{amnon}). Another possible extension is to
consider more general paths between $H_{\init}$ and $H_{\final}$
(see, e.g., \cite{farhipaths, preskill, amnon}).
Obviously, our main results, such as Theorem \ref{thm:main},
hold also for these more general models.

\subsection{Markov Chains and Hermitian Matrices}\label{sec:markovham}
Under certain conditions, there exists a standard mapping of
Hamiltonians to Markov chains (for background on Markov chains, see
\cite{lovasz}).  The following fact is useful to show that this
mapping applies in the case we analyze.

\begin{fact}[Adapted from Perron's Theorem, Theorem $8.2.11$
 in \cite{HornJohnson}]\label{fact:perron}
Let $G$ be a Hermitian matrix with real non-negative entries. If there exists a
finite $k$ such that all entries of $G^k$ are positive, then $G$'s
largest eigenvalue is positive, and all other eigenvalues
are strictly smaller in absolute value.
Moreover, the corresponding eigenvector is unique,
 and all its entries are positive.
\end{fact}

We define the mapping for $G$, a Hermitian matrix operating on an
$L+1$ dimensional Hilbert space.  Suppose that all the entries of $G$
are real and non-negative, that its eigenvector
$(\alpha_0,\ldots,\alpha_L)$ with largest eigenvalue $\mu$ satisfies
$\alpha_i>0$ for all $0\le i\le L$ and that $\mu>0$.  Define $P$ by:
\begin{eqnarray}\label{eq:tran}
P_{ij} &:=& \frac{\alpha_j}{\mu \alpha_i}G_{ij}.
\end{eqnarray}
The matrix $P$ is well defined, and is stochastic because all its entries are
 non-negative and each of its rows sums up to one.  It is easy to
 verify the following fact:
\begin{fact}\label{fact:GH}
The vector $(v_0,\ldots,v_L)$ is an eigenvector of $G$ with eigenvalue $\delta$
if and only if $(\alpha_0 v_0,\ldots,\alpha_L v_L)$ is a left
eigenvector of $P$ with eigenvalue $\delta/\mu$.
\end{fact}

We will consider $G$ of the form $G=I-H$ for
some Hamiltonian $H$.  The above fact implies that if
$(\alpha_0,\ldots,\alpha_L)$ is the ground state of $H$ with
eigenvalue $\lambda$ then $(\alpha_0^2,\ldots,\alpha_L^2)$ is a
left eigenvector of $P$ with maximal eigenvalue $1$. By normalizing,
we obtain that $\pi:=(\alpha_0^2/Z,\ldots,\alpha_L^2/Z)$
is the limiting distribution of $P$,
where $Z=\sum \alpha_i^2$.
Moreover, the gap between
$P$'s largest and second largest eigenvalues is equal to
$\Delta(H)/(1-\lambda)$.

\subsection{Spectral Gaps of Markov Chains}\label{sec:cond}
Given a stochastic matrix $P$ with limiting distribution $\pi$, and a
subset $B\subseteq \{0,\dots,L\}$, the \emph{flow} from $B$ is given
by $F(B) := \sum_{i\in B,j\notin B}{\pi_i P_{ij}}.$ Define the
$\pi$-weight of $B$ as $\pi(B):=\sum_{i\in B}{\pi_i}$.
The \emph{conductance} of $P$ is defined by
$\varphi(P) := \min_{B}{{F(B)}/{\pi(B)}}$,
where we minimize over all non-empty subsets $B\subseteq\{0,\dots,L\}$
with $\pi(B)\leq \smfrac{1}{2}$.

\begin{theorem}[The conductance bound \cite{sinclair})]\label{thm:conductance}
The eigenvalue gap of $P$ is at least $\smfrac{1}{2}\varphi(P)^2$.
\end{theorem}

\section{Equivalence of Adiabatic and Quantum Computation}\label{sec:nogeo}
Here we prove Theorem~\ref{thm:main}, by showing how to simulate a
 quantum circuit with $L$ two-qubit gates on $n$ qubits by an
 adiabatic computation on $n+L$ qubits (the other direction
 was shown in \cite{farhiad, vandamvaz}). We first allow five qubit
 interactions. We later show how to reduce it to three, using
 techniques that will also be used in Section~\ref{sec:geometric}.

\subsection{Five-local Hamiltonian}\label{sec:five}

\begin{theorem}\label{thm:5}
Given a quantum circuit on $n$ qubits with $L$ two-qubit gates
implementing a unitary $U$, and $\epsilon>0$, there exists a $5$-local adiabatic computation
$AC(n+L,2,H_\init,H_\final,\eps)$  whose running time is
$\poly(L,\frac{1}{\epsilon})$ and
whose output (after tracing out some ancilla qubits)
is $\eps$-close (in trace distance) to $U\ket{0^n}$.
Moreover, $H_\init$ and $H_\final$ can be computed by a polynomial
time Turing machine.
\end{theorem}

The running time we obtain here is
$O(\eps^{-(5+3 \delta)}L^{5+2 \delta})$ for any fixed $\delta>0$.

\subsubsection{The Hamiltonian}\label{sec:ham}

For our construction we use the Hamiltonian defined in
\cite{Kitaev:book}.  Denote $\ket{\gamma_\ell}:= \ket{\alpha(\ell)}
\otimes \ket{1^\ell 0^{L-\ell}}^c$, where $\ket{\alpha(\ell)}$ denotes the
state of the circuit after the $\ell$th gate and the superscript $c$ denotes
the clock qubits.  We would like to define a local Hamiltonian $H_{\init}$
with ground state $\ket{\gamma_0}=\ket{0^n}\otimes \ket{0^{L}}^c$, and
a local Hamiltonian
$H_{\final}$ with ground state $\ket{\eta} = \frac{1}{\sqrt{L+1}}
\sum_{\ell=0}^L \ket{\gamma_\ell}$ as in Equation
\ref{eq:final_state}. To do this, we write
$H_{\init}$ and $H_{\final}$ as a sum of terms:
\begin{align*}
H_{\init}&:= H_\clockinit+H_\rminput+H_\clock \\
H_{\final} &:= \frac{1}{2}\sum_{\ell=1}^L{H_\ell}+H_\rminput+H_\clock.
\end{align*}
The terms in $H_{\final}$ (and likewise in $H_{\init}$)
are defined such that the only state whose energy (i.e., eigenvalue)
is $0$ is the desired ground state.  This is done by assigning
an {\em energy penalty} to any state that does not satisfy the required
properties of the ground state.
The different terms, which correspond to different properties
of the ground states, are described in the following paragraphs.
The adiabatic evolution then follows the time-dependent Hamiltonian
\begin{eqnarray} \label{eq:hamham}
H(s)&=&(1-s)H_{\init}+s H_{\final}.
\end{eqnarray}
Notice that as $s$ goes from $0$ to $1$, $H_\clockinit$ is slowly
replaced by $\frac{1}{2}\sum_{\ell=1}^L{H_\ell}$ while
$H_\rminput$ and $H_\clock$ are held constant.

We now describe each of the terms.
First, $H_{\clock}$ checks that the
clock's state is of
the form $\ket{1^\ell 0^{L-\ell}}^c$ for some $0\le \ell\le L$.
This is achieved by assigning an energy penalty to any basis state on the
clock qubits that contains the sequence $01$,
\begin{eqnarray*}
H_{\clock} &:=& \sum_{\ell=1}^{L-1}{\ketbra{01}{01}^c_{\ell,\ell+1}},
\end{eqnarray*}
where the subscript indicates which clock qubits the projection
operates on.  Note that illegal clock states are eigenstates of
$H_{\clock}$ with eigenvalue at least $1$; legal clock states have
eigenvalue $0$.

Next, $H_{\rminput}$ checks that if the clock is $\ket{0^L}^c$,
the computation qubits must be in the state $\ket{0^n}$,
\begin{align*}
H_{\rminput} &:= \sum_{i=1}^n{\ketbra{1}{1}_i}\otimes\ketbra{0}{0}^c_1.
\end{align*}

We complete the description of $H_\init$ with $H_\clockinit$ whose goal is
to check that the clock's state is $\ket{0^L}^c$,
\begin{eqnarray*}
H_{\clockinit}&:=&\ketbra{1}{1}^c_1.
\end{eqnarray*}

\begin{claim}\label{cl:gs}
The state $\ket{\gamma_0}$ is a ground state of $H_{\init}$ with eigenvalue $0$.\footnote{
The state $\ket{\gamma_0}$ is in fact the {\em unique} ground state of $H_{\init}$ as
will become apparent from the proof of the spectral gap. A similar statement
holds for Claim \ref{cl:gs2}.
}
\end{claim}
\begin{proof}
It is easy to verify that $H_\init \ket{\gamma_0} = 0$.
As a sum of projectors, $H_\init$ is positive semidefinite
and hence $\ket{\gamma_0}$ is a ground state of $H_\init$.
\end{proof}

We now proceed to the first term in $H_{\final}$.
The Hamiltonian $H_\ell$ checks that the
propagation from step $\ell-1$ to $\ell$ is correct, i.e., that it
corresponds to the application of the gate $U_\ell$.
For $1 < \ell < L$, it is defined as
\begin{eqnarray} \label{eq:IUUI}
H_\ell &:=&
I\otimes \ketbra{100}{100}^c_{\ell-1,\ell,\ell+1}
- U_\ell\otimes\ketbra{110}{100}^c_{\ell-1,\ell,\ell+1}  \nonumber\\
& & - U_\ell^\dag \otimes \ketbra{100}{110}^c_{\ell-1,\ell,\ell+1}
 + I\otimes\ketbra{110}{110}^c_{\ell-1,\ell,\ell+1}.
\end{eqnarray}
Intuitively, the three-qubit terms above move the state of the clock
one step forward, one step backward, or leave it unchanged.
The accompanying matrices $U_\ell, U^\dag_\ell$ describe the associated time evolution.
For the boundary cases $\ell=1,L$, we omit one
clock qubit from these terms and define
\begin{eqnarray}\label{eq:edge}
H_1  & := &
I\otimes\ketbra{00}{00}_{1,2} -U_1 \otimes\ketbra{10}{00}_{1,2}
 - U_1^\dag \otimes\ketbra{00}{10}_{1,2} + I\otimes \ketbra{10}{10}_{1,2}
\nonumber \\
H_L  & :=&
I\otimes\ketbra{10}{10}_{L-1,L} - U_L \otimes\ketbra{11}{10}_{L-1,L}
  - U_L^\dag \otimes \ketbra{10}{11}_{L-1,L} + I\otimes \ketbra{11}{11}_{L-1,L}.
\end{eqnarray}

\begin{claim}\label{cl:gs2}
The history state $\ket{\eta}$ is a ground state of $H_{\final}$ with eigenvalue $0$.
\end{claim}
\begin{proof}
It is easy to verify that $H_\final \ket{\eta} = 0$.
It remains to notice that for all $1\le \ell\le L$, $H_\ell$ is
positive semidefinite and hence so is $H_\final$.
\end{proof}

\begin{remark}
Strictly speaking, Theorem \ref{thm:5} holds even
if we remove the terms $H_\clock$ and $H_{\rminput}$ from both $H_{\init}$ and $H_{\final}$.
We include them mainly for consistency with the rest of the paper.
\end{remark}

\subsubsection{Spectral Gap in a Subspace}

Let $\calS_0$ be the $L+1$-dimensional subspace spanned by
$|\gamma_0\ra,\ldots,|\gamma_L\ra$.
It is easy to verify the following claim.
\begin{claim}\label{cl:invariant}
The subspace $\calS_0$ is invariant under $H(s)$, i.e., $H(s)(\calS_0)\subseteq \calS_0$
\end{claim}
In this subsection, we show that the spectral gap of $H_{\calS_0}(s)$,
the restriction of $H$ to $\calS_0$, is inverse polynomial in $L$.
As mentioned in Subsection~\ref{sec:model}, this, together with Claim~\ref{cl:invariant},
is enough to obtain a bound on the running time of the adiabatic algorithm.

\begin{lemma}\label{cl:spec}
The spectral gap of the restriction of $H(s)$ to $\calS_0$ satisfies
$\Delta(H_{\calS_0}(s))=\Omega (L^{-2})$ for all $s\in [0,1]$.
\end{lemma}
\begin{proof}
Let us write the Hamiltonians $H_{\calS_0, \init}$ and $H_{\calS_0, \final}$ in the basis
$\ket{\gamma_0},\ldots,\ket{\gamma_L}$ of $\calS_0$.
Both $H_\clock$ and $H_{\rminput}$ are $0$ on  $\calS_0$ and can thus
be ignored.
We have the following $(L+1) \times (L+1)$ matrices:
\begin{eqnarray}\label{eq:h_initial_s0}
H_{\calS_0,\init} & = &
\left(%
\begin{array}{cccc}
  0 &      0 & \ldots      & 0 \\
  0 &      1 & \ldots      & 0 \\
  \vdots & \vdots & \ddots & \vdots \\
  0 &      0 & \ldots & 1  \\
\end{array}%
\right),
\end{eqnarray}
\begin{eqnarray}\label{eq:h_final_s0}
H_{\calS_0,\final} & = & \smfrac{1}{2}\ketbra{\gamma_0}{\gamma_0} - \smfrac{1}{2}\ketbra{\gamma_0}{\gamma_1}
- \smfrac{1}{2}\ketbra{\gamma_{L}}{\gamma_{L\mns1}}+\smfrac{1}{2}\ketbra{\gamma_L}{\gamma_L} \nonumber \\
& &  +  \sum_{\ell=1}^{L-1}({-\smfrac{1}{2}\ketbra{\gamma_\ell}{\gamma_{\ell-1}}+
  \ketbra{\gamma_\ell}{\gamma_\ell} -\smfrac{1}{2}\ketbra{\gamma_\ell}{\gamma_{\ell+1}}}) \nonumber \\
& = & \left(
\begin{array}{rrrrrrr}
\smfrac{1}{2} & \mns \smfrac{1}{2} &0 & & \cdots& & 0 \\ \mns \smfrac{1}{2} & 1 & \mns \smfrac{1}{2} & 0 & \ddots & & \vdots\\ 0 &
\mns \smfrac{1}{2} & 1 & \mns \smfrac{1}{2} & 0 & \ddots & \vdots\\ & \ddots & \ddots & \ddots & \ddots & \ddots & \\ \vdots& &
0 & \mns \smfrac{1}{2} &1 & \mns \smfrac{1}{2}& 0 \\ & & & 0 & \mns \smfrac{1}{2} &1 & \mns \smfrac{1}{2} \\ 0& & \cdots& & 0&
\mns \smfrac{1}{2} & \smfrac{1}{2} \\
\end{array}
\right)
\end{eqnarray}
We now lower bound $\Delta(H_{\calS_0}(s))$. We consider two cases:

\vskip 5pt \noindent{\bf The case $s<1/3$:}
Here, $H_{\calS_0}(s)$ is sufficiently close to $H_{\calS_0, \init}$
(whose spectral gap is $1$) so we can apply the following standard lemma (see, e.g., \cite{bhatia}, Page 244).

\begin{lemma}[Gerschgorin's Circle Theorem] Let $A$ be any matrix with entries $a_{ij}$.
Consider the discs in the complex plane given by
$$ D_i = \Big\{ z ~|~ |z- a_{ii}| \le \sum_{j \neq i} |a_{ij}| \Big\}, ~~ 1 \le i \le n. $$
Then the eigenvalues of $A$ are contained in $\cup D_i$ and any connected
component of $\cup D_i$ contains as many eigenvalues of $A$ as the number of
discs that form this component.
\end{lemma}

For $s < 1/3$, $H_{\calS_0}(s)_{1,1} < 1/6$
 and $\sum_{j \neq 1} H_{\calS_0}(s)_{1,j} < 1/6$.
Moreover, for any $i\neq 1$, $H_{\calS_0}(s)_{i,i} > 5/6$
and $\sum_{j \neq i} H_{\calS_0}(s)_{i,j} < 1/6$. By the above lemma,
we obtain that there is one eigenvalue smaller than $1/3$ while all
other eigenvalues are larger than $2/3$. Hence, the spectral gap is
at least $1/3$.

\vskip 5pt
\noindent{\bf The case $s\ge 1/3$:}
We note that $H_{\calS_0,\final}$ is the Laplacian of the
simple random walk \cite{lovasz} of a particle on a line of length $L+1$.
A standard result in Markov chain theory
implies $\Delta(H_{\calS_0,\final})=\Omega(1/L^2)$ \cite{lovasz}.
For $s \ge 1/3$, $H_{\calS_0}(s)$ is sufficiently close to
$H_{\calS_0, \final}$ to apply Markov chain techniques, as we show next.

Let $(\alpha_0,\ldots,\alpha_L)^\dagger$ be the
ground state of $H_{\calS_0}(s)$ with eigenvalue $\lambda$.
Define the Hermitian matrix $G(s)=I - H_{\calS_0}(s)$.
It is easy to see that $G(s)$ satisfies the conditions of
Fact~\ref{fact:perron} for all $s>0$.
We obtain that the largest eigenvalue $\mu=1-\lambda$
of $G(s)$ is positive and non-degenerate and the corresponding eigenvector
$(\alpha_0,\ldots,\alpha_L)^\dagger$ has positive entries.
We can now map the matrix $G(s)$
 to a stochastic matrix $P(s)$ as described in
 Subsection~\ref{sec:markovham}.
The transition matrix $P(s)$ describes a random walk on the
line of $L+1$ sites (Fig.~\ref{fig:randomwalk}).
Fact \ref{fact:GH} implies that
the limiting distribution of $P(s)$ is given by
 $\pi=(\alpha^2_0/Z,\ldots,\alpha^2_L/Z)$ where $Z=\sum_i \alpha_i^2$.

\begin{figure}[h!]
\center{
 \epsfxsize=3.2in
 \epsfbox{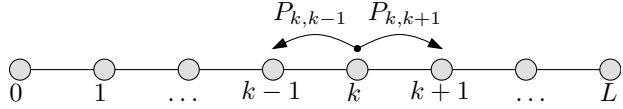}
}
\caption{The random walk of $P(s)$}\label{fig:randomwalk}
\end{figure}

We bound the spectral gap of $P(s)$ using the conductance bound (see
Subsection \ref{sec:cond}).
To do this we need to know that $\pi$ is monotone. We first show:
\begin{claim}\label{claim:mono}
For all $0\le s\le 1$,
the ground state of $H_{\calS_0}(s)$ is monotone, namely
$\alpha_0 \ge \alpha_1\ge  \ldots \ge \alpha_L \ge 0$.
\end{claim}
\begin{proof}
The case $s=0$ is obvious, so assume $s>0$.
We first claim that the ground state $(\alpha_0,\ldots,\alpha_L)^\dagger$
of $H_{\calS_0}(s)=I-G(s)$ can be written as the limit
$$\frac{1}{c_0}\lim_{\ell \rightarrow \infty}(G(s)/\mu)^\ell (1,\dots,1)^\dagger$$ for some constant $c_0>0$.
To see this, let $\ket{v_0},\ldots,\ket{v_L}$ be an orthonormal set of eigenvectors
of $G(s)$, with corresponding
eigenvalues $\mu_0 \ge \mu_1 \ge \ldots \ge \mu_L$. By Fact \ref{fact:perron},
the largest eigenvalue corresponds to a unique eigenvector, and hence we have
$\ket{v_0}=(\alpha_0,\ldots,\alpha_L)^\dagger$, and $\mu_0=\mu$.

The set of eigenvectors $\ket{v_i}$
 forms an orthonormal basis, and we can
write $(1,\ldots,1)^\dagger$ in terms of this basis:
  $(1,\ldots,1)^\dagger=\sum_i c_i \ket{v_i}$.
Now, we have that $(G(s)/\mu)^\ell (1,\dots,1)^\dagger=
\sum_i c_i(\frac{\mu_i}{\mu})^\ell \ket{v_i}$.
By Fact \ref{fact:perron} we have $|\mu_i|<\mu$ for all $i\not=0$, and
$\mu>0$. We thus have that
$\lim_{\ell\rightarrow \infty}(G(s)/\mu)^\ell (1,\dots,1)^\dagger=
 c_0\ket{v_0}$.

It is easy to check that $G(s)$ preserves monotonicity, namely,
if $G(s)$ is applied to a monotone vector, the result is a monotone vector.
Hence, when $G(s)/\mu$ is applied to the monotone vector $(1,\ldots,1)^\dagger$,
the
result is a monotone vector. Thus,
$c_0\ket{v_0}$ is monotone.
Finally, we observe that $c_0> 0$. This is because $c_0$ is the inner product
between the all $1$ vector, and $\ket{v_0}$ whose entries are all
positive by Fact \ref{fact:perron}.  This implies that
$\ket{v_0}$ is also monotone, as desired.
\end{proof}

It follows that $\pi$ is also monotone. We use this and simple
combinatorial arguments to prove
the following claim.
\begin{claim}\label{cl:conduct}
For all $1/3 \le s \le 1$, $\varphi(P(s))\ge \smfrac{1}{6 L}$.
\end{claim}

\begin{proof}
We show that for any nonempty $B \subseteq \{0,\ldots,L\}$,
$F(B)/\pi(B) \ge \smfrac{1}{6L}$. We consider two cases.
First, assume that $0\in B$. Let $k$ be the smallest
such that $k\in B$ but $k+1 \notin B$. Then,
$$F(B)\ge \pi_k P(s)_{k,k+1}= \pi_k \cdot \frac{\sqrt{\pi_{k+1}}}{\mu \sqrt{\pi_k}} G(s)_{k,k+1}=
  \frac{\sqrt{\pi_k \pi_{k+1}}}{1-\lambda} G(s)_{k,k+1} \ge \frac{\pi_{k+1}}{1-\lambda} G(s)_{k,k+1}$$
where the last inequality follows from the monotonicity of $\pi$. Using the definition of $G$
and the assumption that $s \ge 1/3$ we get
that $G(s)_{k,k+1}\geq {1/6}$. We also have
$0<1-\lambda \le 1$, where the second inequality follows from the fact that
 $H_{\calS_0}(s)$ is positive semidefinite, and the first follows from
$\mu>0$ which we previously deduced from Fact \ref{fact:perron}.
 Hence,
\begin{align}\label{eq:kcase1}
 \frac{F(B)}{\pi(B)} \ge \frac{\pi_{k+1}}{6\pi(B)}
\end{align}
By $\pi(B) \le 1/2$, we have $\pi(\{k+1,\ldots,L\}) \ge 1/2$.
Together with $\pi(\{k+1,\ldots,L\}) \le L\pi_{k+1}$ we obtain
$\pi_{k+1} \ge 1/(2L)$. This yields the desired bound
$F(B)/\pi(B) \ge 1/(6L)$.

Now assume that $0\notin B$ and let $k$ be the smallest
such that $k \notin B$ and $k+1 \in B$.
It is easy to see that $\pi_k P(s)_{k,k+1} = \pi_{k+1} P(s)_{k+1,k}$.
Hence, using the same argument as before we can see that
Equation \ref{eq:kcase1} holds in this case too.
Since $B \subseteq \{k+1,\ldots,L\}$, we have $\pi(\{k+1,\ldots,L\}) \ge
\pi(B)$. Hence, $\pi_{k+1} \ge \pi(B)/L$. Again, this yields the bound
$F(B)/\pi(B) \ge 1/(6L)$.
\end{proof}
By Theorem~\ref{thm:conductance}, we have that the spectral gap
of $P(s)$ is larger than $1/(2\cdot (6)^2 \cdot L^2)$.
By Subsection~\ref{sec:markovham}, we have that $\Delta(H_{\calS_0})\ge
\mu/(2\cdot (6)^2 L^2)$. Finally, notice that $\mu = 1 - \lambda
\ge \frac{1}{2}$, because $\lambda\le
\bra{\gamma_0}H_{\calS_0}(s)\ket{\gamma_0}=\frac{s}{2}\le \frac{1}{2}$.
\end{proof}

\subsubsection{Running Time}\label{sec:nonsense}

We now complete the proof of Theorem \ref{thm:5}.
Note that we have already proved something which is very close
to Theorem \ref{thm:5}.
\begin{claim}\label{cl:lookwhatwehavedone}
Given a quantum circuit on $n$ qubits with $L$ gates,
the adiabatic algorithm with $H_\init$ and $H_\final$ as
defined in the previous section,
with
$T=O(\eps^{-\delta}L^{4+2 \delta})$ for some fixed $\delta>0$,
outputs a final state that is within $\ell_2$-distance $\eps$ of
the history state of the circuit, $\ket{\eta}$.
The running time of the algorithm is $O(T\cdot L)$.
\end{claim}
\begin{proof}
Claim~\ref{cl:invariant} shows that $\calS_0$ is invariant under $H$.
Hence, as mentioned in Subsection~\ref{sec:model},
an adiabatic evolution according to $H$
is identical to an adiabatic evolution according to $H_{\calS_0}$.
Using Lemma \ref{cl:spec} and Theorem \ref{thm:ad} (with $\|H_\init-H_\final\|=O(1)$),
we obtain that for $T$ as above the final state (with global phase adjusted appropriately) is indeed $\eps$-close in $\ell_2$-norm to $\ket{\eta}$.
By our definition, the running time of the adiabatic algorithm
is $O(T\cdot L)$ since $\|H(s)\|\le (1-s)\|H_{\init}\|+s\|H_{\final}\|=O(L+n)=O(L)$.
The last equality follows from $n=O(L)$, because each qubit is assumed
to participate in the computation (otherwise we can omit it).
\end{proof}

In fact, one might be satisfied with this claim, which enables
generating adiabatically a state
which is very close to $\ket{\eta}$, instead of our desired
 $\ket{\alpha(L)}$.
To see why this might be sufficient to simulate quantum circuits,
suppose for a moment that $\eps$ is $0$, and the final
state is exactly $\ket{\eta}$.
As mentioned in the introduction, we can now
measure the clock qubits of
the history state, and with probability $1/L$ the outcome is $\ell=L$, which
means that the state of the first register
is the desired state $\ket{\alpha(L)}$.
If the measurement yields another value, we repeat the adiabatic algorithm
from scratch.
To get $\ell=L$ with sufficiently high probability,
we repeat the process $O(L)$ times, which introduces an overhead factor
of $L$. The above discussion
is also true with $\eps>0$, as long as it is much smaller than
$1/L$, the weight of $\ket{\alpha(L)}$ in $\ket{\eta}$.

However, this is not sufficient
 to complete the proof of Theorem~\ref{thm:5}.
Indeed, the theorem as stated follows our definition of the
model of adiabatic computation, which
allows to perform one adiabatic evolution and then measure (and possibly trace out some qubits).
Classical postprocessing such as conditioning on $\ell$ being equal to
$L$, and repeating the computation if it is not, are not allowed.
Hence, we need to adiabatically generate a state that is close to
$\ket{\alpha(L)}$.

This technical issue can be resolved with the following
simple trick, which at the same time allows us to avoid the overhead factor
of $L$ introduced before. We simply
add another $O(\frac{1}{\eps}L)$
identity gates to the original quantum circuit at the end of
its computation and then apply the adiabatic simulation
to this modified circuit.
This modification increases the weight of $\ket{\alpha(L)}$ in the
history state.
The following lemma makes this precise.
\begin{lemma}\label{lem:out}
Assume we can transform any given quantum circuit with $L$ two-qubit
gates on $n$ qubits into a $k$-local adiabatic
computation on n+L $d$-dimensional particles whose output is $\epsilon$ close in $\ell_2$-norm to
the history state of the quantum circuit and whose running
time is $f(L,\epsilon)$ for some function $f$.
Then, we can transform any given quantum circuit
with $L$ two-qubit gates on $n$ qubits into a $k$-local
adiabatic computation on $n+2L/\eps$ $d$-dimensional particles
 whose output (after
tracing out some ancilla qubits) is $\epsilon$ close in trace
 distance to the final state of
the circuit and whose running time is $f(2L/\eps, \eps/2)$.
\end{lemma}
\begin{proof}
Given a quantum circuit on $n$ qubits with $L$ gates,
consider the circuit obtained by appending to it $(\frac{2}{\epsilon}-1)L$
identity gates. Let $L' = 2L/\epsilon$ be the number of
gates in the modified circuit and let
$\ket{\eta}$ denote its history state. By our assumption,
we can transform this modified circuit into an adiabatic
computation whose output
is $\eps/2$ close in $\ell_2$-norm to $\ket{\eta}$ and whose
 running time is $f(L', \eps/2)$.
Since the trace distance between two pure states is bounded from
above by the $\ell_2$-distance (see, e.g.,
\cite{computingmixedstates}), we obtain that the output of the
adiabatic computation
is also $\eps/2$ close in trace distance to $\ketbra{\eta}{\eta}$.
In addition, it is easy to check that after
we trace out the clock qubits from $\ket{\eta}$, we are left with a
state that is $\eps/2$ close
in trace distance to the final state of the circuit. We complete
the proof by applying the triangle inequality.
\end{proof}

We can now apply this lemma on the result of
Claim \ref{cl:lookwhatwehavedone}.
This completes the proof
of Theorem \ref{thm:5}, with the running time being
$O(\eps^{-(5+3\delta)}L^{5+2 \delta})$.

\subsubsection{Spectral Gap}\label{ssec:spectralgap}

In the previous subsections, we presented a Hamiltonian $H(s)$ and showed that
inside a preserved subspace $\calS_0$ it has a non-negligible spectral gap.
This was enough for the adiabatic algorithm since the entire
adiabatic evolution is performed inside this subspace.

In this subsection, we show that the spectral gap of $H(s)$
in the entire Hilbert space is also non-negligible.
The purpose of this result is twofold. First,
the existence of a non-negligible spectral gap
in the entire Hilbert space
might have some relevance when dealing with adiabatic
computation in the presence of noise (see, e.g., \cite{preskill}).
Second, the techniques that we use here, and in particular Lemma \ref{lem:gapinsides},
are used again in the next subsection, but are easier to present in this
simpler context.

\begin{lemma}
For all $0\le s\le 1$, $\Delta(H(s))=\Omega(L^{-3})$.
\end{lemma}
\begin{proof}
Let $\calS$ be the subspace of dimension $(L+1) \cdot 2^n$ spanned by all legal clock states.
Observe that $\calS$ is preserved by $H(s)$, i.e., $H(s)(\calS)\subseteq \calS$.
Hence, the eigenstates of $H(s)$ belong either to $\calS$ or to
its orthogonal subspace $\calS^\perp$. We can therefore analyze the spectrum
of $H_\calS(s)$ and of $H_{\calS^\perp}(s)$ separately.

First, due to the term $H_{\clock}$ and the fact that all
other terms are positive semidefinite, the ground energy
of $H_{\calS^\perp}(s)$ is at least $1$. Second, as we will show next using
Lemma~\ref{lem:gapinsides}, the spectral gap of $H_\calS(s)$
is $\Omega(L^{-3})$. To establish the same spectral gap for $H(s)$,
it is enough to show that the ground energy of $H_{\calS}(s)$
is smaller than $\frac{1}{2}$, which would mean that the spectral
gap of $H(s)$ is exactly that of $H_{\calS}(s)$. Indeed,
 observe that
$$\bra{\gamma_0} H_{\calS}(s) \ket{\gamma_0}= \bra{\gamma_0} H_{\calS_0}(s) \ket{\gamma_0}
  = s/2 \le 1/2$$
where the first equality holds because $\ket{\gamma_0} \in \calS_0$ and the second
follows from Equations \ref{eq:h_initial_s0} and \ref{eq:h_final_s0}.
Therefore, the smallest eigenvalue of $H_{\calS}(s)$ is bounded from above
by $1/2$.
\end{proof}

\begin{lemma}\label{lem:gapinsides}
Let $\calS$ denote the subspace spanned by all legal clock states.
Then the ground state of $H_{\calS}(0)$ is $\ket{\gamma_0}$, and that
of $H_{\calS}(1)$ is $\ket{\eta}$.
Moreover, for all $0\le s\le 1$, $\Delta(H_\calS(s))=\Omega(L^{-3})$.
\end{lemma}

\begin{proof}
We can write $\calS$ as the direct sum of $2^n$ orthogonal
subspaces $\calS_0,\calS_1,\ldots,\calS_{2^n-1}$, defined as follows.
For $0\le j\le 2^n-1$ and $0\le \ell\le L$ define
$\ket{\gamma^j_\ell}:=\ket{\alpha^j(\ell)} \otimes \ket{1^\ell 0^{L-\ell}}$,
where $\ket{\alpha^j(\ell)}$ is the state of the quantum circuit at time $\ell$ if
the input state corresponds to the binary representation $j$.
Note that $\ket{\gamma^0_\ell}=\ket{\gamma_\ell}$.
The space $\calS_j$ is spanned by $\{\ket{\gamma^j_0},\ldots,\ket{\gamma^j_L}\}$.
It is easy to check the following claim (see Figure~\ref{fig:block}).
\begin{claim}\label{cl:blocks}
The Hamiltonian $H_\calS(s)$ is block diagonal in the $\calS_j$'s.
\end{claim}
\begin{figure}[h!]
\center{
 \epsfxsize=2in
 \epsfbox{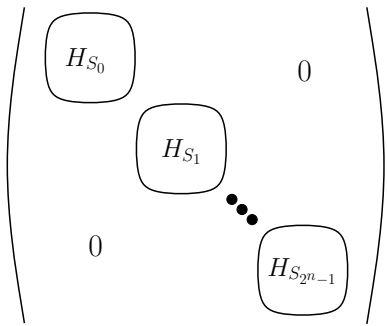}
}
\caption{$H_{\calS}(s)$ is block diagonal.}
\label{fig:block}
\end{figure}

By Claims~\ref{cl:gs}, \ref{cl:gs2}, \ref{cl:spec}, and \ref{cl:blocks},
it suffices to argue that the ground energy of $H_{\calS_j}(s)$
for any $j\ne 0$ is larger than the ground energy of
$H_{\calS_0}(s)$ by at least $\Omega(1/L^3)$.
Essentially, this follows from the penalty given by the term $H_{\rminput}$
to nonzero input states. The proof, however, is
slightly subtle since $H_{\rminput}$ assigns a penalty only to
states $\ket{\gamma_\ell^j}$ with $\ell=0$.

Notice that
$$H_{\calS_j}(s)=H_{\calS_0}(s)+H_{\calS_j,\rminput}.$$
Moreover, for $1\le j\le 2^n-1$, $H_{\calS_j,\rminput}$ is diagonal, with
its top-left element at least $1$ (it actually equals
the number of $1$'s in the binary representation of $j$)
and all other diagonal elements zero. Hence, if we define
$M$ as
\begin{eqnarray*}
M &:=& \left(%
\begin{array}{cccc}
  1 &      0 & \ldots      & 0 \\
  0 &      0 & \ldots      & 0 \\
  \vdots & \vdots & \ddots & \vdots \\
  0 &      0 & \ldots & 0  \\
\end{array}%
\right)
\end{eqnarray*}
then $H_{\calS_j,\rminput}-M$ is positive definite and therefore
we can lower bound the ground energy of $H_{\calS_j}(s)$ with the ground
energy of $H_{\calS_0}(s)+M$.
For this, we apply the following geometrical lemma by Kitaev (Lemma 14.4 in
\cite{Kitaev:book}).

\begin{lemma}
Let $H_1,H_2$ be two Hamiltonians with
ground energies $a_1,a_2$, respectively.
Suppose that for both Hamiltonians the difference between the energy of the (possibly degenerate) ground space and the next highest eigenvalue is
larger than $\Lambda$,
and that the angle between the two ground spaces is $\theta$. Then
the ground energy of $H_1+H_2$ is at least
 $a_1+a_2+2\Lambda \sin^2(\theta/2)$.
\end{lemma}
 \dnote{define angle between subspaces}
We now apply this lemma to $H_{\calS_0}(s)$ and $M$. By Claim~\ref{cl:spec}, the spectral gap of $H_{\calS_0}(s)$ is
$\Omega(1/L^2)$. The spectral gap of $M$ is clearly $1$. Moreover, using Claim \ref{claim:mono}, we obtain that the
angle between the two ground spaces satisfies $\cos(\theta)\le 1-1/L$ by the monotonicity property of the ground state
of $H_{\calS_0}(s)$ (see Claim \ref{claim:mono}). It follows that the ground energy of $H_{\calS_j}(s)$ is higher by at
least $\Omega(1/L^3)$ than that of $H_{\calS_0}(s)$.
\end{proof}

\begin{remark}
Notice that we only used the following properties of $H_\rminput$: its restriction
to $\calS_0$ is $0$ and its restriction to $\calS_j$ for any $j\neq 0$ is a diagonal
matrix in the basis $\ket{\gamma^j_0},\ldots,\ket{\gamma^j_L}$ whose top-left entry is at least $1$ and all other entries are non-negative.
This observation will be useful in Section \ref{sec:geometric}.
\end{remark}

\subsection{Three-local Hamiltonian}\label{sec:three}
We now show that adiabatic computation with $3$-local Hamiltonians
is sufficient to simulate standard quantum computations.
\begin{theorem}\label{thm:3}
Given a quantum circuit on $n$ qubits with $L$ two-qubit gates implementing
a unitary $U$, and $\eps>0$, there exists
a $3$-local adiabatic computation $AC(n+L,2,H_\init,H_\final,\eps)$
whose running time is
$\poly(L,\frac{1}{\epsilon})$ and whose output state is
$\eps$-close (in trace distance) to $U \ket{0^n}$.
Moreover, $H_\init$ and $H_\final$ can be computed by a polynomial
time Turing machine.
\end{theorem}

The proof of this theorem builds on techniques developed in previous
subsections.
The techniques developed in this section will be used in
Section~\ref{sec:geometric}.

\subsubsection{The Hamiltonian}

Consider the Hamiltonian constructed in Subsection \ref{sec:ham}.
Notice that all terms except $H_\ell$ are already $3$-local (some are even $2$-local or $1$-local).
In order to obtain a $3$-local Hamiltonian, we remove two clock qubits from the $5$-local
terms in $H_\ell$ and leave only the $\ell$th clock qubit. More precisely,
for $1< \ell < L$ define
\begin{eqnarray*}
 H'_\ell &:=&
I\otimes \ketbra{100}{100}^c_{\ell-1,\ell,\ell+1}- U_\ell\otimes\ketbra{1}{0}^c_{\ell}
- U_\ell^\dag \otimes \ketbra{0}{1}^c_{\ell}
+ I\otimes\ketbra{110}{110}^c_{\ell-1, \ell,\ell+1}.
\end{eqnarray*}
For the boundary cases $l=1,L$ we define
\begin{align*}
 H'_1 &:=
I\otimes \ketbra{00}{00}^c_{1,2}- U_1\otimes\ketbra{1}{0}^c_{1}
- U_1^\dag \otimes \ketbra{0}{1}^c_{1}
+ I\otimes\ketbra{10}{10}^c_{1,2} \\
 H'_L &:=
I\otimes \ketbra{10}{10}^c_{L-1,L}- U_\ell\otimes\ketbra{1}{0}^c_{L}
- U_L^\dag \otimes \ketbra{0}{1}^c_{L}
+ I\otimes\ketbra{11}{11}^c_{L-1,L}.
\end{align*}

Note that because of the terms $\ketbra{1}{0}^c$ and $\ketbra{0}{1}^c$,
these Hamiltonians no longer leave the subspace $\calS$ invariant.
To mend this, we assign a much larger
energy penalty to illegal clock states. As we will see soon, this
makes the lower part of the spectrum of our Hamiltonians behave essentially
like in their restriction to $\calS$.
Set  $J=\eps^{-2} L^6$ and define\onote{this can also be taken to be $c \eps^{-2} L^5$ for some large
enough constant}
\begin{eqnarray*}
H'_{\init}&:= & H_{\clockinit} + H_{\rminput}+J \cdot H_\clock \\
H'_{\final} &:= &\frac{1}{2}\sum_{\ell=1}^L{H'_\ell}+H_{\rminput}+J \cdot H_\clock.
\end{eqnarray*}
The Hamiltonian we use here is thus
$$H'(s)=(1-s)H'_{\init}+sH'_{\final}.$$
Essentially the same proof as that of Claim \ref{cl:gs} shows that $\ket{\gamma_0}$ is a ground state of $H'_{\init}$.
However, it turns out that $\ket{\eta}$ is no longer a ground state of $H'_{\final}$ (the proof of Claim \ref{cl:gs2}
does not apply since $H'_\ell$ is no longer positive semidefinite). However, as we shall see later, $\ket{\eta}$ is
very close to the ground state of $H'_{\final}$.

\subsubsection{The Spectral Gap}\label{sec:gapthree}

Our first claim is that, when restricted to $\calS$, $H'$ and
$H$ are identical.

\begin{claim}\label{clm:h_hprime_same}
For any $0\le s\le 1$, $H_\calS(s) = H'_\calS(s)$.
\end{claim}
\begin{proof}
Let $\Pi_\calS$ be the orthogonal projection on $\calS$. Then our goal is
to show that $\Pi_\calS H(s) \Pi_\calS = \Pi_\calS H'(s) \Pi_\calS$.
The only difference between $H(s)$ and $H'(s)$ is the factor of $J$ in $H_\clock$, and that the $H_\ell$ terms are
replaced by $H'_\ell$. We note that $H_{\calS,\clock}$ is zero.
Hence, it suffices to show that for all $1 \le \ell \le L$,
$$ \Pi_\calS H_\ell \Pi_\calS = \Pi_\calS H'_\ell \Pi_\calS.$$
For this, observe that for any $1 < \ell < L$,
$$ \Pi_\calS  \ketbra{1}{0}^c_{\ell} \Pi_\calS =
   \ketbra{1^\ell 0^{L-\ell}}{1^{\ell-1} 0^{L-(\ell-1)}}^c
 = \Pi_\calS  \ketbra{110}{100}^c_{\ell-1,\ell,\ell+1} \Pi_\calS$$
and similarly for $\ketbra{0}{1}^c_{\ell}$. A similar statement holds for $\ell=1,L$
with the right hand term modified appropriately.
\end{proof}

 Lemma~\ref{lem:gapinsides} and Claim~\ref{clm:h_hprime_same}
imply that $\Delta(H'_\calS(s))=\Omega(L^{-3})$.
We now want to deduce from this a lower bound on
 $\Delta(H'(s))$, without the restriction to $\calS$.
 For this we use the following claim.
Essentially, it says that if $J$ is large enough,
then the lower part of the spectrum of $H'(s)$ is
similar to that of $H'_\calS(s)$. More precisely, it shows
that the lowest eigenvalues, the second lowest eigenvalues,
and the ground states of the two Hamiltonians are close.
Intuitively, this holds since the energy penalty given to
states in $\calS^\perp$, the orthogonal space to $\calS$, is very high
and hence any eigenvector with low eigenvalue must be almost orthogonal
to $\calS^\perp$ (and hence almost inside $\calS$).
We note that a similar lemma was used in
\cite{Kempe:03a} in the context of quantum $\NP$-complete problems.

\begin{lemma}\label{le:leak}
Let $H=H_1+H_2$ be the sum of two Hamiltonians operating on some
Hilbert space $\calH=\cal S+\cal S^\perp$. The Hamiltonian $H_2$ is
such that $\cal S$ is a zero eigenspace and the eigenvectors in $\cal
S^\perp$ have eigenvalue at least $J> 2K$ where $K=\|H_1\|$. Let $a$
and $b$ be the lowest and the second lowest eigenvalues of $H_\calS$
and let $a'$ and $b'$ be the corresponding quantities for $H$. Then
the lowest eigenvalue of $H$ satisfies $a - \frac{K^2}{J-2K} \le a'
\le a$ and the second lowest eigenvalue of $H$ satisfies $b'\ge b-
\frac{K^2}{J-2K}$. If, moreover, $b>a$ then the ground states
$\ket{\xi},\ket{\xi'}$ of $H_\calS,H$ respectively satisfy
\begin{eqnarray*}
 |\langle \xi | \xi' \rangle | ^2
 & \ge& 1- \frac{K^2}{(b-a)(J-2K)}.
\end{eqnarray*}
\end{lemma}
\begin{proof}
First, we show that $a'\le a$. Using $H_2|\xi \ra = 0$,
\begin{eqnarray*}
\bra{\xi}H \ket{\xi}&=&\bra{\xi}H_1 \ket{\xi}+\bra{\xi}H_2 \ket{\xi}
~=~ a
\end{eqnarray*}
and hence $H$ must have an eigenvector of eigenvalue at most $a$.

We now show the lower bound on $a'$. We can write any unit
vector $\ket{v} \in \calH$ as $\ket{v}=\alpha_1
\ket{v_1}+\alpha_2 \ket{v_2}$ where $\ket{v_1} \in \cal S$ and $\ket{v_2} \in \calS^\perp$ are two unit vectors and
$\alpha_1,\alpha_2$ are two non-negative reals satisfying $\alpha_1^2+\alpha_2^2=1$. Then we have,
\begin{eqnarray*}\label{eq:projection}
\bra{v} H \ket{v}
&\geq& \bra{v} H_1 \ket{v} + J \alpha_2^2 \nonumber \\
&=& (1-\alpha_2^2) \bra{v_1}H_1 \ket{v_1} + 2 \alpha_1 \alpha_2 {\rm Re} \bra{v_1} H_1
               \ket{v_2}+\alpha_2^2 \bra{v_2} H_1 \ket{v_2} + J \alpha_2^2 \nonumber \\
&\geq& \bra{v_1}H_1 \ket{v_1} - K \alpha_2^2 - 2 K \alpha_2
                - K \alpha_2^2 + J \alpha_2^2 \nonumber \\
&=& \bra{v_1}H_1 \ket{v_1} + (J-2K) \alpha_2^2 - 2 K \alpha_2
\end{eqnarray*}
where we used $\alpha_1^2 = 1-\alpha_2^2$ and $\alpha_1 \le 1$. Since $(J-2K) \alpha_2^2 - 2 K \alpha_2$ is minimized
for $\alpha_2=K/(J-2K)$, we have
\begin{eqnarray}\label{eq:projection2}
\bra{v} H \ket{v} &\geq& \bra{v_1}H_1 \ket{v_1} - \frac{K^2}{J-2K}.
\end{eqnarray}
We obtain the required lower bound by noting that $\bra{v_1}H_1 \ket{v_1} \ge a$.

Consider now the two-dimensional space ${\cal L}$ spanned by the two eigenvectors of $H$ corresponding to $a'$ and
$b'$. For any unit vector $\ket{v} \in {\cal L}$ we have $\bra{v}H\ket{v} \le b'$. Hence, if ${\cal L}$ contains a
vector $\ket{v}$ orthogonal to $\calS$, then we have $b' \ge \bra{v}H\ket{v} \ge J-K > K \ge b$ and we are done.
Otherwise, the projection of ${\cal L}$ on $\calS$ must be a two-dimensional space. Being two-dimensional, this space
must contain a vector orthogonal to $\ket{\xi}$.
 Let $\ket{v}$ be a vector in ${\cal L}$ whose projection on $\calS$
is orthogonal to $\ket{\xi}$. By \eqref{eq:projection2},
 $b' \ge \bra{v}H\ket{v} \ge b - \frac{K^2}{J-2K}$, as
required.

Finally, let $\beta = |\langle \xi | \xi' \rangle | ^2$.
Then we can write $\ket{\xi} =
\sqrt{\beta}\ket{\xi'}+\sqrt{1-\beta}\ket{\xi'^\perp}$ for some
unit vector $\ket{\xi'^\perp}$ orthogonal to
$\ket{\xi'}$. Since $\ket{\xi'}$ is an eigenvector of $H$,
\begin{eqnarray*}
a ~=~ \bra{\xi}H\ket{\xi} &=& \beta \bra{\xi'}H\ket{\xi'} +
(1-\beta)\bra{\xi'^\perp}H\ket{\xi'^\perp}\\
  & \ge &\beta a' + (1-\beta) b' \\
    &\ge& \beta \Big(a - \frac{K^2}{J-2K}\Big) +
(1-\beta)\Big(b - \frac{K^2}{J-2K}\Big) \\
    &=  &a + (1-\beta)(b-a) - \frac{K^2}{J-2K}.
\end{eqnarray*}
Rearranging, we obtain the required bound.
\end{proof}

We can now bound the spectral gap of $H'(s)$.
 \begin{lemma}\label{lm:spec3}
For all $0\le s\le 1$, $\Delta(H'(s))=\Omega(L^{-3})$.
\end{lemma}
\begin{proof}
We apply Lemma~\ref{le:leak} by setting $H_2=J\cdot H_{clock}$ and $H_1$ to be the remaining terms such that
$H'(s)=H_1+H_2$. Note that Lemma~\ref{le:leak} implies that the spectral gap of $H'(s)$ is smaller than that of
$H'_{\calS}(s)$ (which is $\Omega(1/L^3)$ by Lemma~\ref{lem:gapinsides}) by at most $K^2/(J-2K)$. But it is easy to see
that $K=O(L)$, due to the fact that $H_1$ consists of $O(L)$ terms, each of constant norm.
The result follows since $J =\eps^{-2}L^6$.
\end{proof}
This shows the desired bound on the spectral gap.
Before we complete the proof, we must show that the final ground state
is close to the history state.
\begin{lemma}\label{lm:gs_close}
The ground state of
$H'(1)$ is $\eps$-close to $\ket{\eta}$.
\end{lemma}
\begin{proof}
Apply Lemma~\ref{le:leak} as in the proof
of Lemma \ref{lm:spec3}, for the case
$s=1$. We obtain that the inner product squared between the ground state of $H'(1)$ and $\ket{\eta}$, is at least
$1-\delta$, with $\delta= \frac{K^2}{(b-a)(J-2K)}=O(L^{-1}\eps^2)$,
 where we have used $K=O(L)$, $J=\eps^{-2} L^{6}$, and $b - a
=\Omega(1/L^3)$ by Lemma~\ref{lem:gapinsides}.
This implies that the $\ell_2$-distance between the ground state of
$H'(1)$ and $\ket{\eta}$ is $O(\eps/\sqrt{L}) \leq \eps$.
\end{proof}

We now complete the proof of Theorem~\ref{thm:3}.
The adiabatic algorithm starts with $\ket{\gamma_0}$ and evolves according
to $H'(s)$ for  $T=\theta(\eps^{-\delta}L^{7+3\delta})$.
Such a $T$ satisfies the adiabatic condition
(Equation \ref{eq:adiabatic_cond}), using $\|H'_{\final}-H'_{\init}\|=O(L)$.
By Theorem \ref{thm:ad} the final state is
$\eps$-close in $\ell_2$-distance to the ground state of $H'_\final$.
Lemma \ref{lm:gs_close} implies that this state is $\eps$-close in
$\ell_2$-distance to $\ket{\eta}$.
 Using the triangle inequality we note that the output of the
adiabatic computation is $2\eps$-close to $\ket{\eta}$.
The running time of this algorithm is $O(T \cdot J \cdot L)=O(T\cdot \eps^{-2}L^7)=
O(\eps^{-(2+\delta)}L^{14+3\delta})$.

 We can now apply Lemma \ref{lem:out} to obtain a modified
adiabatic computation whose output state after
tracing out the clock qubits is $\eps$-close in trace
distance to $U|0^n\ra$. The running time is
$O(\eps^{-(16+4\delta)}L^{14+3 \delta})$ for any fixed $\delta>0$.

%%%%%%%%%%%%%%%%%%%%%%%%%%%%%%%%%%%%%%%%%%%%%%%%%%%%%%%%%%%%%%%%%%%%%
\section{Two Local Hamiltonians on a Two-Dimensional Lattice}\label{sec:geometric}
%%%%%%%%%%%%%%%%%%%%%%%%%%%%%%%%%%%%%%%%%%%%%%%%%%%%%%%%%%%%%%%%%%%%%

In this section we prove Theorem~\ref{thm:geo}.
We simulate a given quantum circuit by an adiabatic evolution of a system
of $6$-dimensional quantum particles arranged on a two-dimensional grid.
More precisely, we prove the following theorem:
\begin{theorem}\label{thm:6level}
Given a quantum circuit on $n$ qubits with $L$ two-qubit gates
implementing a unitary $U$, and $\epsilon>0$, there exists a $2$-local adiabatic computation
$AC(\poly(n,L),6,H_\init,H_\final,\eps)$ such that $H_\init$
and $H_\final$ involve only nearest neighbors on a $2$-dimensional grid,
its running time is $\poly(L,\frac{1}{\epsilon})$,
and its output  (after performing a partial measurement on each particle)
 is $\eps$-close (in trace distance) to
$U \ket{0^n}$. Moreover,
$H_\init$ and $H_\final$ can be computed by a polynomial time Turing
machine.
\end{theorem}

As mentioned in the introduction, the main problem
 in proving this theorem, and more precisely,  in
 moving to a two dimensional grid, is the notion of a clock.
In the constructions of the previous section,
the clock is represented by an additional
register that counts the clock steps in unary
representation.
The terms $H_\ell$, which check the correct propagation in the $\ell$th time step,
interact between the $\ell$th qubit of the clock and
 the corresponding qubits on which $U_\ell$ operates. If we want
to restrict the interaction to nearest neighbors in two dimensions using this
idea, then no matter how the clock qubits are arranged on the grid, we run into problems interacting the qubits with the corresponding clock qubits in a local way.
The solution to this problem lies in the way we represent the clock.
Instead of using an extra register, we embed the clock
into the same particles that perform the computation
by defining the notion of a {\em shape} of a state, to be defined later.
We then create a sequence of legal shapes, and show how states can evolve
from one legal shape to another.

Although the construction of this section is more involved than the ones of
the previous section, its analysis follows almost immediately from the analysis
carried out in Theorem \ref{thm:3}. To achieve this, we make sure that
the Hamiltonians and some relevant subspaces are as similar as possible
to those in the previous section.

\subsection{Assumptions on the Input Circuit}\label{sec:layout}

\begin{figure}[ht]
\center{ \epsfxsize=5in\epsfbox{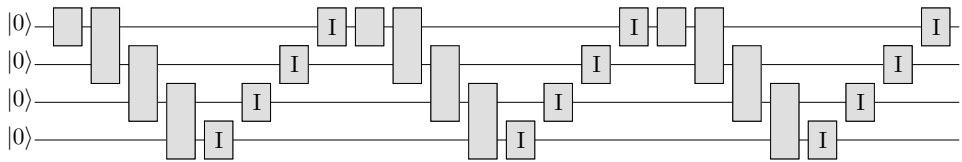}}
\caption{The modified circuit with $R=3$.}
\label{figure:circuit}
\end{figure}

To simplify the construction of our adiabatic evolution, we first
assume without loss of generality that
the quantum circuit we wish to simulate has a particular layout of its
gates.
Namely, it consists of $R$ rounds, where each round is
composed of $n$ nearest neighbor
gates (some can be the identity gate), followed by $n$ identity gates, as in Figure \ref{figure:circuit}.
More specifically, the first gate in each round is a one-qubit gate applied to the
first qubit. For $i=2,\dots,n$, the $i$th gate is a two-qubit gate
applied to qubits $i-1$ and $i$. For $i=n+1, \ldots, 2n$ the $i$th gate is an identity gate applied to
the $(2n+1-i)$th qubit. These identity gates are included for convenience of notation.
Any circuit can be transformed to such a form by introducing extra identity
and swap gates. Let $L=2nR$ be the total number of gates in the circuit so obtained.
Clearly, $L$ is at most polynomially larger than the number of gates in the original circuit.

\subsection{The Particles of the Adiabatic Quantum System}
The adiabatic computation is performed on $6$-dimensional particles,
arranged on a two-dimensional square lattice with $n$ rows and $R+1$
columns. We number the rows from $1$ (top) to $n$ (bottom) and
the columns from $0$ (left) to $R$ (right).
Columns number $0$ and $1$ are used to simulate the first round of the circuit.
Columns number $1$ and $2$ are used for the second round of computation,
and so on. We denote the six internal states of a particle by $\ket{\statea},
\ket{\stateba},\ket{\statebb},
\ket{\stateca},\ket{\statecb}$, and $\ket{\stated}$.
These six states are divided into
four {\em phases}: the \emph{unborn phase} $\ket{\statea}$,
the {\em first phase} $\ket{\stateba},\ket{\statebb}$, the {\em second phase}
$\ket{\stateca},\ket{\statecb}$, and the {\em dead phase} $\ket{\stated}$.
The two states in the first phase and
the two states in the second phase correspond
to computational degrees of freedom, namely to
the $\ket{0}$ and $\ket{1}$ states of a qubit. We write
$\ket{\stateb}$ to denote an arbitrary state in the subspace
spanned by $\ket{\stateba}$ and $\ket{\statebb}$.
Similarly,  $\ket{\statec}$ denotes a state in the space
spanned by $\ket{\stateca}$ and $\ket{\statecb}$.
The phases are used to define the {\em shape} of the basis states.
A shape of a basis state is simply an assignment of
one of the four phases to each particle, ignoring the
computational degrees of freedom inside the first and second phase.
These shapes will be used instead of the clock states of the
previous section.

\subsection{Geometrical Clock}
We now describe the way we represent clock using shapes.  In the previous
constructions, the space $\calS$ of dimension $2^n(L+1)$ was the
ground space of the clock, i.e., the space spanned by legal clock
states. Inside the clock register there were $L+1$ legal clock states.
Note that each such clock state can be
described, essentially, in a geometric way by the ``shape'' of the clock
particles: how many $1$'s precede how many $0$'s.

\begin{figure}[h!]
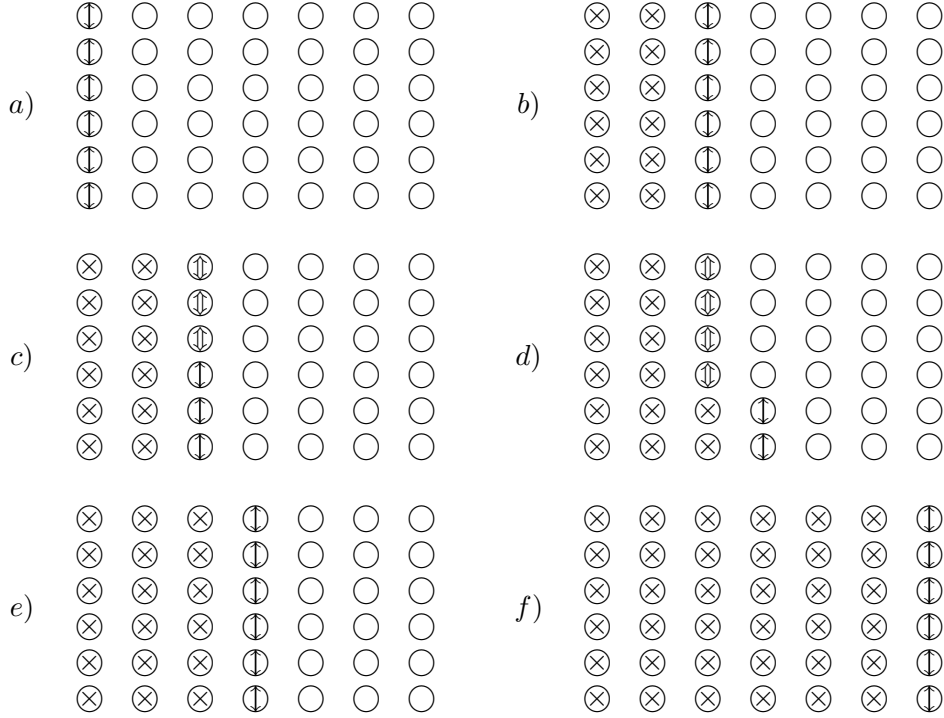

\center{{
%{\setlength{\arraycolsep}{0.5mm}
\begin{equation*}
\begin{array}{cccc}
a) &
\begin{array}{ccccccc}
\stateb & \statea & \statea & \statea & \statea & \statea & \statea \\
\stateb & \statea & \statea & \statea & \statea & \statea & \statea \\
\stateb & \statea & \statea & \statea & \statea & \statea & \statea \\
\stateb & \statea & \statea & \statea & \statea & \statea & \statea \\
\stateb & \statea & \statea & \statea & \statea & \statea & \statea \\
\stateb & \statea & \statea & \statea & \statea & \statea & \statea
\end{array}&~~~~b) &
\begin{array}{ccccccc}
\stated & \stated & \stateb & \statea & \statea & \statea & \statea \\
\stated & \stated & \stateb & \statea & \statea & \statea & \statea \\
\stated & \stated & \stateb & \statea & \statea & \statea & \statea \\
\stated & \stated & \stateb & \statea & \statea & \statea & \statea \\
\stated & \stated & \stateb & \statea & \statea & \statea & \statea \\
\stated & \stated & \stateb & \statea & \statea & \statea & \statea
\end{array}\\\nonumber
&&&\\\nonumber
c) &
\begin{array}{ccccccc}
\stated & \stated & \statec & \statea & \statea & \statea & \statea \\
\stated & \stated & \statec & \statea & \statea & \statea & \statea \\
\stated & \stated & \statec & \statea & \statea & \statea & \statea \\
\stated & \stated & \stateb & \statea & \statea & \statea & \statea \\
\stated & \stated & \stateb & \statea & \statea & \statea & \statea \\
\stated & \stated & \stateb & \statea & \statea & \statea & \statea
\end{array}&~~~~d)&
\begin{array}{ccccccc}
 \stated & \stated & \statec & \statea & \statea & \statea & \statea \\
 \stated & \stated & \statec & \statea & \statea & \statea & \statea \\
 \stated & \stated & \statec & \statea & \statea & \statea & \statea \\
 \stated & \stated & \statec & \statea & \statea & \statea & \statea \\
 \stated & \stated & \stated & \stateb & \statea & \statea & \statea \\
 \stated & \stated & \stated & \stateb & \statea & \statea & \statea
\end{array}\\\nonumber
&&&\\\nonumber
e)&
\begin{array}{ccccccc}
\stated & \stated & \stated & \stateb & \statea & \statea & \statea \\
\stated & \stated & \stated & \stateb & \statea & \statea & \statea \\
\stated & \stated & \stated & \stateb & \statea & \statea & \statea \\
\stated & \stated & \stated & \stateb & \statea & \statea & \statea \\
\stated & \stated & \stated & \stateb & \statea & \statea & \statea \\
\stated & \stated & \stated & \stateb & \statea & \statea & \statea
\end{array}&~~~~f)&
\begin{array}{ccccccc}
 \stated & \stated & \stated & \stated & \stated & \stated & \stateb \\
 \stated & \stated & \stated & \stated & \stated & \stated & \stateb \\
 \stated & \stated & \stated & \stated & \stated & \stated & \stateb \\
 \stated & \stated & \stated & \stated & \stated & \stated & \stateb \\
 \stated & \stated & \stated & \stated & \stated & \stated & \stateb \\
 \stated & \stated & \stated & \stated & \stated & \stated & \stateb
\end{array}
\end{array}\nonumber
\end{equation*}
%}
}}
\caption{Legal clock states}
\label{figure:states}
\end{figure}

We now describe the corresponding
subspaces involved in our construction for the
two dimensional case.
For each $0\le \ell \le L$, we have a $2^n$-dimensional subspace
corresponding to that clock state. Each of these $L+1$ subspaces
can be described
by its {\em shape}, that is, a setting of one of the four phases to
each particle.
%As before, we will have $L+1$ legal shapes,
%where the $j$th shape corresponds to the $j$th
%step of the circuit.
%As in previous sections $\calS$ is the direct sum of these subspaces.
%It is the space spanned by all states with legal shapes.
See Figure~\ref{figure:states} for an illustration
with $n = 6, R=6$. The six shapes shown correspond to clock states
$\ell=0, \ell=4n, \ell=4n+3, \ell=5n+2, \ell=6n$,
 and $\ell=2nR$ respectively.  Notice that each shape
 has exactly $n$ particles in the first or
second phase. Hence, the dimension of the subspace induced
by each shape is $2^n$.
As $\ell$ goes from $0$ to $L$, the shape changes from that
shown in Fig.~\ref{figure:states}a to that shown
in Fig.~\ref{figure:states}. The locations at which
the changes occur form a snake-like pattern winding
down and up the lattice, following the layout of the gates in the input circuit Fig.~\ref{figure:circuit}.

We now describe the legal shapes more formally.
\begin{enumerate} \item The shape
corresponding to clock state
$\ell=2nr+k$ for $0\le k\le n$ has its $r$ leftmost columns
in the dead phase. The top $k$ particles in the $r+1$st column
are in their second phase while the bottom $n-k$ are in the
first phase. Particles in the remaining $R-r$ columns are all
in the unborn phase.
\item  The shape corresponding to clock state
$\ell=2nr+n+k$ for $1\le k\le n-1$ has, as before, its
$r$ leftmost columns in the dead phase. The $r+1$st column
has its $n-k$ topmost particles in the second phase,
and its remaining $k$ particles in the dead phase.
The $r+2$nd column has its $n-k$ topmost particles in the
unborn phase and its remaining $k$ particles in the
first phase. All remaining particles are in the unborn
phase.
\end{enumerate}

\begin{figure}[h!]
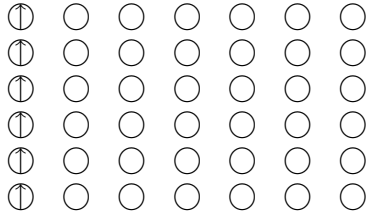

\center{{
%{\setlength{\arraycolsep}{0.5mm}
\begin{equation*}
\begin{array}{ccccccc}
\stateba & \statea & \statea & \statea & \statea & \statea & \statea \\
\stateba & \statea & \statea & \statea & \statea & \statea & \statea \\
\stateba & \statea & \statea & \statea & \statea & \statea & \statea \\
\stateba & \statea & \statea & \statea & \statea & \statea & \statea \\
\stateba & \statea & \statea & \statea & \statea & \statea & \statea \\
\stateba & \statea & \statea & \statea & \statea & \statea & \statea
\end{array}
\end{equation*}
}}
\caption{The initial state}
\label{figure:firststate}
\end{figure}

The subspace $\calS$ is defined as the $(L+1)2^n$-dimensional
space spanned by all legal shapes.
As in previous sections we partition $\calS$ into $2^n$ subspaces $\calS_j$.
Each subspace $\calS_j$ is spanned by $L+1$ orthogonal
states $\ket{\gamma^j_0},\ldots,\ket{\gamma^j_L}$, defined as
follows. For each $0\le \ell\le L$ and $0\le j\le 2^n-1$,
the shape of $\ket{\gamma^j_\ell}$ corresponds to $\ell$.
The state of the $n$ active particles (i.e., those in
either the first or the second phase), when read from top to bottom,
corresponds to the state of the circuit after the first $\ell$ gates are
applied to an initial state corresponding to the binary representation
of $j$; i.e., it corresponds to the state $U_\ell \cdot U_{\ell-1}\cdots U_1|j\ra$.
More precisely, these particles are in a superposition obtained by
mapping this state to the state of the $n$ active particles
in the following way:  $\ket{0}$ to $\stateba$
(or $\stateca$ for a second phase particle)
and $\ket{1}$ to $\statebb$ (or $\statecb$ for a second phase particle).
We often denote $\ket{\gamma_\ell^0}$, which corresponds to the
all $0$ input, by $\ket{\gamma_\ell}$.
For example, $\ket{\gamma_0}$ is shown in
Figure \ref{figure:firststate}.

With the risk of being somewhat redundant, let us now give
an alternative description of the states
$\ket{\gamma_0^j},\ldots,\ket{\gamma_L^j}$. This description
is more helpful in understanding the Hamiltonians $H''_\ell$ which
we will define shortly.
Consider a state $\ket{\gamma^j_\ell}$ for some $\ell=2rn$.
The $n$ particles in the $r$th column are in their first phase
and their computational degrees of freedom correspond to the
state of the circuit's qubits at the beginning of the $r$th round.
Particles to the left of this column are dead, those to the right
of this column are unborn. The state $\ket{\gamma^j_{\ell+1}}$
is obtained from $\ket{\gamma^j_{\ell}}$ by changing
the topmost particle in the $r$th column to a second phase
particle and applying the first gate in the $r$th round (a one-qubit gate)
to its computational degrees of freedom.
Next, the state $\ket{\gamma^j_{\ell+2}}$
is obtained from $\ket{\gamma^j_{\ell+1}}$ by changing
the second particle from above in the $r$th column to a second phase
particle and applying the second gate in the $r$th round (a two-qubit gate)
to both this particle and the one on top of it.
We continue in a similar fashion until we reach
$\ket{\gamma^j_{\ell+n}}$, in which the entire
$r$th column is in the second phase.
We refer to these steps as the downward stage.

Next, let us describe the upward stage.
The state $\ket{\gamma^j_{\ell+n+1}}$ is obtained from $\ket{\gamma^j_{\ell+n}}$
by `moving' the bottommost particle in the $r$th column one location
to the right. More precisely, the bottommost particle changes to the dead phase
and the one to the right of it changes to the first phase. The computational
degrees of freedom are the same in both states. This corresponds to the
fact that the $n+1$st gate in a round of the circuit is the identity gate.\footnote{We
could allow arbitrary one-qubit gates here instead of identity gates.
This leads to a slightly more efficient construction but also to more cumbersome
Hamiltonians.}
Continuing in a similar fashion, we see that the upwards stage ends in the
state $\ket{\gamma^j_{\ell+n+n}} = \ket{\gamma^j_{2(r+1)n}}$
that matches the above description of the first state in a round.

\subsection{The Hamiltonian}
The initial and final Hamiltonians are defined as
\begin{eqnarray*}
H''_{\init} &:=& H''_{\clockinit}
~+~ H''_{\rminput} ~+~ J \cdot H''_{\clock}\\
H''_{\final} &:=& \frac{1}{2}\sum_{\ell=1}^L{H''_\ell}
+ H''_{\rminput} +
J \cdot H''_{\clock},
\end{eqnarray*}
where $J=\eps^{-2}\cdot L^6$.
These Hamiltonians are chosen to be as similar as possible to the
corresponding Hamiltonians in previous sections.
For example, $H''_\clock$ has as its ground space
the space of legal clock states, $\calS$. As before, it
allows us to essentially project all other Hamiltonians
on $\calS$, by assigning a large energy penalty to states with illegal shape.
Also, the Hamiltonians $H''_\ell$ (once projected to $\calS$)
check correct propagation from one step to the next.
Other terms also serve similar roles as before.

Let us start with the simplest terms.
Define
$$H''_{\rminput} := \sum_{i=1}^{n}{(\ketbra{\statebb}{\statebb})_{i,1}}.$$
The indices indicate the row and column of the particle on which
the Hamiltonian operates.
This Hamiltonian checks that none of the particles in
the leftmost column are in $\ket{\statebb}$.
Then, define
$$H''_{\clockinit}=(I-\ketbra{\stateba}{\stateba}-\ketbra{\statebb}{\statebb})_{1,1}.$$
This Hamiltonian checks that the top-left particle
 is in a $\ket{\stateb}$ state.
The remaining terms are described in the following subsections.

\subsubsection{The Clock Hamiltonian}
The shapes we define satisfy the following important property:
there exists a {\em two-local way} to verify that a shape
is legal.
This allows us to define a two-local clock Hamiltonian,
 $H''_\clock$, whose ground space is
exactly $\calS$, the $(L+1)2^n$-dimensional
space spanned by all legal shapes.
%This is done using the local rules.

\begin{table}[h!]
\center{
\begin{tabular}{|l|l|}
\hline
Forbidden  & Guarantees that \\ \hline \hline
   $\statea\stateb, \statea\statec, \statea\stated$& $\statea$ is to the right of all other qubits\\ \hline
   $\statea\stated, \stateb\stated, \statec\stated$& $\stated$ is to the left of all other qubits\\ \hline
   $\statea\stated, \stated\statea$& $\statea$ and $\stated$ are not horizontally adjacent\\ \hline
   $\stateb\stateb$, $\stateb\statec$,&\\
 $\statec\stateb$, $\statec\statec$ & only one of $\stateb$, $\statec$ per row \\ \hline
   $\ontop{\statea}{\statec}, \ontop{\stateb}{\statec}, \ontop{\stated}{\statec}$& only $\statec$ above $\statec$\\ \hline
   $\ontop{\stateb}{\statea}, \ontop{\stateb}{\statec}, \ontop{\stateb}\stated$& only $\stateb$ below $\stateb$\\ \hline
   $\ontop{\statea}{\stated}, \ontop{\stated}{\statea}$& $\statea$ and $\stated$ are not vertically adjacent\\ \hline
   $\ontop{\statec}{\statea}, \ontop{\stated}{\stateb}$& no $\statea$ below $\statec$ and no $\stateb$ below $\stated$\\ \hline
\end{tabular}
}
\caption{Local rules for basis state to be in $\calS$}
\label{tab:rules}
\end{table}

\begin{claim}\label{cl:rules}
A shape is legal if and only if it contains none of the forbidden configurations of Table~\ref{tab:rules}.
\end{claim}
\begin{proof}
It is easy to check that any legal shape contains none of the forbidden configurations.
For the other direction, consider any shape that contains none of these configurations.
Observe that each row must be of the form $\stated^*[\stateb,\statec]\statea^*$, that is,
it starts with a sequence of zero or more $\stated$, it then contains either
$\stateb$ or $\statec$, and then ends with a sequence of zero or more $\statea$.
Columns can be of three different forms. Read from top to bottom, it is
either $\statec^*\stateb^*$, $\statec^*\stated^*$, or $\statea^*\stateb^*$.
It is now easy to verify that such a shape must be one of the legal shapes.
\end{proof}

Using this claim, we can define a two-local
nearest-neighbor Hamiltonian that guarantees a legal shape.
For example, if the rule forbids a particle at location $(i,j)$
in state $\statea$ to the left of a particle at location $(i,j+1)$ in state $\stated$,
then the corresponding term in the Hamiltonian is
$(\ketbra{\statea,\stated}{\statea,\stated})_{(i,j),(i,j+1)}$.
Summing over all the forbidden configurations
 of Table \ref{tab:rules} and over all relevant pairs of particles, we have
$$H''_{\clock} := \sum_{r\in \rules} H_{r}.$$
Note that the ground space of $H''_\clock$ is the $(L+1)2^n$-dimensional space $\calS$.

\subsubsection{The Propagation Hamiltonian}
The choice of legal shapes has the following important
property: the shape of $\ell$
and that of $\ell+1$ differ in at most two locations. This means that
 for any $\ell$ and $j$, the shape of
$\ket{\gamma_{\ell-1}^j}$ and that of $\ket{\gamma_\ell^j}$
differ in at most two locations. Moreover, if we consider
the state of the $n$ active particles in both states
we see that these differ on at most two particles, namely,
those on which the $\ell$th gate in the circuit acts.
Crucially, and this is where we use our assumption
on the form of the circuit (Figure \ref{figure:circuit}),
the particle(s) on which the $\ell$th gate acts are at the same
location as the particle(s) whose phase changes. It is
this structure that allows us to define the
Hamiltonians $H''_\ell$. These Hamiltonians act on two
particles and `simultaneously' advance the clock
(by changing the shape) and advance the computational
state (by modifying the state of the active particles).
%The terms $H''_\ell$, $1\le \ell\le L$ are defined
% to check the correct propagation from $\ket{\gamma_{\ell-1}}$
%to $\ket{\gamma_{\ell}}$.
Since $\ket{\gamma_\ell}$ differs from
$\ket{\gamma_{\ell-1}}$ in at most two adjacent lattice sites, this can
be done using a two-body nearest neighbor Hamiltonian.

The definition of $H''_\ell$ depends on whether $\ell$ is in the
downward phase (i.e., is of the form $2rn+k$ for $1\le k\le n$)
or in the upward phase (i.e., is of the form $2rn+n+k$ for $1\le k\le n$).
\onote{the following needs to be checked whenever the style of the paper
changes  to make sure the lines are still the same}
We first define $H''_\ell$ for the upward phase. Assume $\ell = 2rn+n+k$ for
some $0\le r < R,1< k< n$ and let $i=n-k+1$ be the row in
which $\ket{\gamma_{\ell-1}}$ and $\ket{\gamma_\ell}$ differ. Then,
\begin{eqnarray*}
H''_\ell &:=&
 \stackketbra{\stateca}{\stated} \hskip -5pt \ontop{\rm_{i,r}}{\rm_{i+1,r}}
 +  \stackketbra{\statea}{\stateba} \hskip -5pt \ontop{\rm_{i-1,r+1}}{\rm_{i,r+1}}
 -  \left(\ketbra{\stateca,\statea}{\stated,\stateba} +  \ketbra{\stated,\stateba}{\stateca,\statea}\right)_{\rm(i,r)(i,r+1)}
 \\
& + &
 \stackketbra{\statecb}{\stated} \hskip -5pt \ontop{\rm_{i,r}}{\rm _{i+1,r}} +
 \stackketbra{\statea}{\statebb} \hskip -5pt \ontop{\rm_{i-1,r+1}}{\rm_{i,r+1}}
- \left(\ketbra{\statecb,\statea}{\stated,\statebb} + \ketbra{\stated,\statebb}{\statecb,\statea}\right)_{\rm(i,r)(i,r+1)}.
\end{eqnarray*}
The first line corresponds to changing the state $\ket{\stateca,\statea}$ into
$\ket{\stated,\stateba}$.
 The second line is similar for
$\ket{\statecb,\statea}$ and $\ket{\stated,\statebb}$.
The purpose of the first two terms in each line is the same as
that of
 $\ketbra{100}{100}^c$ and $\ketbra{110}{110}^c$ in $H_\ell$ from
previous sections.\footnote{There are other (equally good) ways
to define these terms. For example, it is possible to define them
so that they both act on the $r$th column.}
The difference is that here, to
uniquely identify the current clock state, we need to consider
particles on top of each other. The remaining terms in each line
correspond to $\ketbra{100}{110}^c$ and $\ketbra{100}{110}^c$ in $H_\ell$

For the case $k=1,n$, the definition is
\begin{eqnarray*}
H''_{2rn+n+1} &:=&
 \ketbra{\stateca}{\stateca}_{n,r}
 +  \stackketbra{\statea}{\stateba} \hskip -5pt \ontop{\rm_{n-1,r+1}}{\rm_{n,r+1}}
 -  \left(\ketbra{\stateca,\statea}{\stated,\stateba} +  \ketbra{\stated,\stateba}{\stateca,\statea}\right)_{\rm(n,r)(n,r+1)}
 \\
& + &
 \ketbra{\statecb}{\statecb}_{n,r} +
 \stackketbra{\statea}{\statebb} \hskip -5pt \ontop{\rm_{n-1,r+1}}{\rm_{n,r+1}}
- \left(\ketbra{\statecb,\statea}{\stated,\statebb} + \ketbra{\stated,\statebb}{\statecb,\statea}\right)_{\rm(n,r)(n,r+1)}\\
H''_{2rn+2n} &:=&
 \stackketbra{\stateca}{\stated} \hskip -5pt \ontop{\rm_{1,r}}{\rm_{2,r}}
 +  \ketbra{\stateba}{\stateba}_{1,r+1}
 -  \left(\ketbra{\stateca,\statea}{\stated,\stateba} +  \ketbra{\stated,\stateba}{\stateca,\statea}\right)_{\rm(1,r)(1,r+1)}
 \\
& + &
 \stackketbra{\statecb}{\stated} \hskip -5pt \ontop{\rm_{1,r}}{\rm _{2,r}} +
 \ketbra{\statebb}{\statebb}_{1,r+1}
- \left(\ketbra{\statecb,\statea}{\stated,\statebb} + \ketbra{\stated,\statebb}{\statecb,\statea}\right)_{\rm(1,r)(1,r+1)}.
\end{eqnarray*}

For the downward stage, $H''_\ell$ checks that a gate is
applied correctly. For $\ell=2nr+k$ and $1<k<n$ we define
\begin{align*}
H''_\ell := \left(\begin{array}{cc} 0& -U_\ell\\ -U_\ell^{\dagger}&0
\end{array}\right) &+
\left(
\stackketbra{\stateca}{\stateba} +
\stackketbra{\stateca}{\statebb} +
\stackketbra{\statecb}{\stateba} +
\stackketbra{\statecb}{\statebb}
\right)\ontop{\rm_{k-1,r}}{\rm_{k,r}} \\
&+ \left(
\stackketbra{\stateca}{\stateba} +
\stackketbra{\stateca}{\statebb} +
\stackketbra{\statecb}{\stateba} +
\stackketbra{\statecb}{\statebb}
\right)\ontop{\rm_{k,r}}{\rm_{k+1,r}}.
\end{align*}
The last two terms are meant, as before, to
replace the terms $\ketbra{110}{110}^c$ and $\ketbra{100}{100}^c$.
Once again, to uniquely identify the current clock state, we need
to consider particles on top of each other.
The first term represents a Hamiltonian that acts on the two
particles in positions $(k,r)$ and $(k+1,r)$. These
particles span a $36$-dimensional space. The matrix shown
above is in fact the restriction of this Hamiltonian to the $8$
dimensional space spanned by
$$ \stackket{\stateca}{\stateba} ~~ \stackket{\stateca}{\statebb} ~~ \stackket{\statecb}{\stateba} ~~ \stackket{\statecb}{\statebb}~~~~
   \stackket{\stateca}{\stateca} ~~ \stackket{\stateca}{\statecb} ~~ \stackket{\statecb}{\stateca} ~~ \stackket{\statecb}{\statecb}$$
(recall that $U_\ell$ acts on two qubits and is therefore a $4\times 4$ matrix). Everywhere else in this $36$ dimensional subspace,
this Hamiltonian acts trivially, i.e., is 0.

For the case $k=n$ we slightly modify the terms that identify the clock states,
\begin{align*}
H''_{2nr+n} := \left(\begin{array}{cc} 0& -U_{2nr+n}\\ -U_{2nr+n}^{\dagger}&0
\end{array}\right) &+
\left(
\stackketbra{\stateca}{\stateba} +
\stackketbra{\stateca}{\statebb} +
\stackketbra{\statecb}{\stateba} +
\stackketbra{\statecb}{\statebb}
\right)\ontop{\rm_{n-1,r}}{\rm_{n,r}} \\
&+ \left(
\ketbra{\stateca}{\stateca} +
\ketbra{\statecb}{\statecb}
\right)_{n,r}.
\end{align*}
For the case $k=1$ we have
\begin{align*}
H''_{2nr+1} := \left(\begin{array}{cc} 0& -U_{2nr+1}\\ -U_{2nr+1}^{\dagger}&0
\end{array}\right) &+
\left(
\ketbra{\stateba}{\stateba} +
\ketbra{\statebb}{\statebb}
\right)_{1,r} \\
&+ \left(
\stackketbra{\stateca}{\stateba} +
\stackketbra{\stateca}{\statebb} +
\stackketbra{\statecb}{\stateba} +
\stackketbra{\statecb}{\statebb}
\right)\ontop{\rm_{1,r}}{\rm_{2,r}},
\end{align*}
where the first term shows the restriction an operator acting on the particle $(1,r)$
to the four dimensional space spanned by $\ket{\stateba}, \ket{\statebb}, \ket{\stateca}, \ket{\statecb}$
(recall that $U_{2nr+1}$ is
a one-qubit gate).

\subsection{Spectral Gap}\label{sec:spectralgap6}

The analysis of the spectral gap follows almost immediately from that in Subsection
\ref{sec:gapthree}. The main effort is in verifying that
the restriction of each of our Hamiltonians to $\calS$ is identical to
the restriction of the corresponding Hamiltonian
in previous sections to $\calS$, when both are constructed
according to the modified quantum circuit of
Subsection~\ref{sec:layout}.
This, in fact, does not hold for $H''_{\rminput}$,
whose projection is not quite the same as that of $H_\rminput$;
still, it is similar enough for the analysis in Subsection
\ref{sec:gapthree} to hold.

\begin{claim}
 $H''_{\calS, \clockinit} = H_{\calS, \clockinit}$
\end{claim}
\begin{proof}
 Both Hamiltonians are diagonal
in the basis $\ket{\gamma_\ell^j}$ with eigenvalue
$0$ for $\ell=0$ and eigenvalue $1$ for any $\ell>0$.
\end{proof}

\begin{claim}
For any $1\le \ell \le L$, $H''_{\calS, \ell}=H_{\calS, \ell}$.
\end{claim}
\begin{proof}
It is straightforward to verify that
both Hamiltonians, when restricted to $\calS$, are equal to
$$\sum_{j=0}^{2^n-1}[\ketbra{\gamma^j_\ell}{\gamma^j_\ell}+\ketbra{\gamma^j_{\ell-1}}{\gamma^j_{\ell-1}}-
\ketbra{\gamma^j_\ell}{\gamma^j_{\ell-1}}-\ketbra{\gamma^j_{\ell-1}}{\gamma^j_\ell}].$$
\end{proof}

For $H''_{\rminput}$ the situation is similar,
although in this case the restriction to $\calS$ is not exactly the same.
Still, the resemblance is enough for the same analysis to hold:

\begin{claim}
Both $H_{\calS,\rminput}$ and
 $H''_{\calS, \rminput}$ are diagonal in the basis $\ket{\gamma_\ell^j}$.
Moreover, the eigenvalue in both Hamiltonians
 corresponding to $\ket{\gamma_\ell^j}$
for $\ell=0$ is exactly the number of $1$'s in the binary
 representation of $j$.
\end{claim}

\begin{proof} Easy to verify.
\end{proof}
The similarity between the two Hamiltonians breaks down as follows.
While the eigenvalues corresponding to $\ket{\gamma_\ell^j}$
 for $\ell>0$
are $0$ in $H_{\calS, \rminput}$, those in $H''_{\calS, \rminput}$ might be positive
(namely, for $0\le \ell \le n$, the eigenvalue of $\ket{\gamma_\ell^j}$ is
the number of $1$'s in the last $n-\ell$ digits in the binary representation
of $j$). Nevertheless,  due to the remark at the end of Subsection
 \ref{ssec:spectralgap},
Lemma \ref{lem:gapinsides} holds here as well.
We get:

\begin{lemma}
For any $0\le s\le 1$, $H''_\calS(s)$ has a spectral
gap of $\Omega(L^{-3})$.
Moreover,
 the ground state of $H''_{\calS, \final}$ is $\ket{\eta}$.
\end{lemma}

The rest of the proof of Theorem \ref{thm:geo}
 is essentially the same as in Subsection \ref{sec:gapthree}. By applying Lemma \ref{le:leak}, we obtain that
\begin{lemma}
For all $0\le s \le 1, \Delta(H''(s))=\Omega(L^{-3})$.
Moreover, the ground state of $H''(1)$ is $\epsilon$-close to $\ket{\eta}$.
\end{lemma}
The proof is similar to that of Lemmas \ref{lm:spec3} and \ref{lm:gs_close}.  This enables us to adiabatically generate the history state with exactly
the same running time as in the three-local case (when the number
of gates is that of the modified circuit of Subsection \ref{sec:layout}).

Finally, we would like to apply
Lemma \ref{lem:out} as before.
However, we cannot quite do this due to a technical issue:
our Hilbert space is no longer a tensor product of
computation qubits and clock qubits
and tracing out the clock qubits is meaningless.
Nevertheless, a minor modification of that lemma still applies.
We first add, say, $L/\epsilon$
identity gates to the end of the (modified) circuit.
Now, the adiabatic computation produces a state close to the
history state. We then measure the shape of the system
without measuring the inner computational degrees of freedom.
Due to the additional identity gates, with all but $\eps$ probability,
the outcome of the measurement is a shape
$\ell$ for $\ell \ge L$. If this is the case then the state of the system
is such that the active particles are in the final state
of the circuit, as desired.
This completes the proof of the theorem.

\section{Acknowledgments}
 We wish to thank Dave Bacon, Ed Farhi, Leonid Gurvitz and Umesh
 Vazirani for inspiring discussions.
DA's work is supported in part by
 ARO grant DAAD19-03-1-0082, NSF ITR grant CCR-0121555,
 ISF grant 032-9738 and an Alon fellowship.
WvD work was supported in part by the
U.S.\ Department of Energy (DOE) and cooperative research
agreement DF-FC02-94ER40818, a CMI postdoctoral fellowship,
and an  HP/MSRI fellowship.
JK's effort is partly sponsored by DARPA and AFL,
 Air Force Material Command, USAF,
 under agreement number F30602-01-2-0524 and FDN00014-01-1-0826 and by ACI S\'ecurit\'e Informatique,
2003-n24, projet ``R\'eseaux Quantiques", ACI-CR 2002-40 and EU 5th
framework program RESQ IST-2001-37559.
OR's effort is supported by an Alon Fellowship, the Israeli Science Foundation, ARO grant DAAD19-03-1-0082,
 and NSF grant CCR-9987845.
Part of this work was done while DA, WvD, JK, ZL were
 members of the MSRI, Berkeley, CA.

%\bibliographystyle{ieee}
%\bibliography{adbib}
%\begin{thebibliography}{99}

\end{document}